\documentclass[11pt,document,nofootinbib,superscriptaddress,onecolumn,preprintnumbers,balancelastpage]{article}
\pdfoutput=1
\usepackage{jheppub}
\usepackage{graphicx}
\usepackage{epstopdf}
\usepackage{dcolumn}% Align table columns on decimal point
\usepackage{bm}% bold math
\usepackage{hyperref}
\usepackage{booktabs}
\usepackage[dvipsnames]{xcolor}
\usepackage{amsmath}
\usepackage{cancel}
\usepackage{xpatch}
\usepackage{mathtools}
\definecolor{c1}{rgb}{0.368417, 0.506779, 0.709798}
\definecolor{c2}{rgb}{0.880722, 0.611041, 0.142051}
\definecolor{c3}{rgb}{0.560181, 0.691569, 0.194885}
\definecolor{c4}{rgb}{0.922526, 0.385626, 0.209179}
\definecolor{c5}{rgb}{0.528488, 0.470624, 0.701351}
\definecolor{c6}{rgb}{0.772079, 0.431554, 0.102387}
\definecolor{c7}{rgb}{0.363898, 0.618501, 0.782349}
\definecolor{turq}{rgb}{0.181,0.638,0.594}
\definecolor{pink}{rgb}{1.000,0.54,0.8}
\definecolor{purple}{RGB}{155,100,155}
\definecolor{gray}{RGB}{128,128,128}
\definecolor{lightBlue}{RGB}{148,179,229}
\definecolor{lightRed}{RGB}{213,157,131}
\definecolor{violet}{RGB}{130,121,173}
\definecolor{gold}{RGB}{255,191,0}

\def\beq{\begin{align}}
\def\eeq{\end{align}}

\newcommand{\vev}[1]{ \left\langle {#1} \right\rangle }

\def\Mpl{M_{\rm Pl}}
\def\GeV{{\rm GeV}}

\title{
Dark Matter Detection, Standard Model Parameters, and
Intermediate Scale Supersymmetry
}

\author[1,2]{David Dunsky}
\author[1,2]{Lawrence J. Hall}
\author[3]{Keisuke Harigaya}
\affiliation[1]{Department of Physics, University of California, Berkeley, California 94720, USA}
\affiliation[2]{Theoretical Physics Group, Lawrence Berkeley National Laboratory, Berkeley, California 94720, USA}
\affiliation[3]{School of Natural Sciences, Institute for Advanced Study, Princeton, New Jersey 08540, USA}

\abstract{
The vanishing of the Higgs quartic coupling at a high energy scale may be explained by Intermediate Scale Supersymmetry, where supersymmetry breaks at $(10^9$-$10^{12})$ GeV. 
The possible range of supersymmetry breaking scales can be narrowed down by precise measurements of the top quark mass and the strong coupling constant. On the other hand, nuclear recoil experiments can probe Higgsino or sneutrino dark matter up to a mass of $10^{12}$ GeV. We derive the correlation between the dark matter mass and precision measurements of standard model parameters, including supersymmetric threshold corrections.
The dark matter mass is bounded from above as a function of the top quark mass and the strong coupling constant. The top quark mass and the strong coupling constant are bounded from above and below respectively for a given dark matter mass.
We also discuss how the observed dark matter abundance can be explained by freeze-out or freeze-in during a matter-dominated era after inflation, with the inflaton condensate being dissipated by thermal effects.
}

\date{\today}

\begin{document}
\maketitle
\flushbottom

%\tableofcontents
\newpage

\section{Introduction}

In 1985, Goodman and Witten proposed that halo dark matter could be detected directly in terrestrial experiments by observing small energy depositions from elastic scattering of dark matter particles from nuclei \cite{Goodman:1984dc}. Their first illustration was of a neutral particle, such as a heavy neutrino, scattering via $t$-channel $Z$ exchange with a cross section per nucleon of order $\sigma v \sim G_F^2 \mu_{\rm red}^2 / 2 \pi$, where $\mu_{\rm red}$ is the reduced mass of the dark matter and nucleon. They computed a signal of order $10^2 - 10^4$ events per Kg per day for dark matter masses in the GeV to TeV range, depending on nuclear target.  In the intervening 35 years, a succession of ever larger and more sensitive detectors have excluded this example by many orders of magnitude, so that the focus has shifted to theories where there is no contribution to the scattering from tree-level weak interactions. In fact, as the number density of dark matter particles scales as the inverse of its mass, present data constrains the mass of dark matter with tree-level $Z$ exchange to be larger than $3 \times 10^9$ GeV~\cite{Aprile:2018dbl}. Proposed detectors \cite{Aprile:2015uzo,Aalbers:2016jon,Akerib:2018lyp} will probe the mass range 
\begin{align}
    %M_{EWDM}
    M_{{\rm DM},Z\mathchar`-{\rm exchange}}
    = (3 \times 10^9 - 2 \times 10^{12}) \; {\rm GeV}.
    \label{eq:MDMEW}
\end{align}

\begin{figure}[tb]
    \centering
    \begin{minipage}{0.5\textwidth}
        \centering
        \includegraphics[width=.95\textwidth]{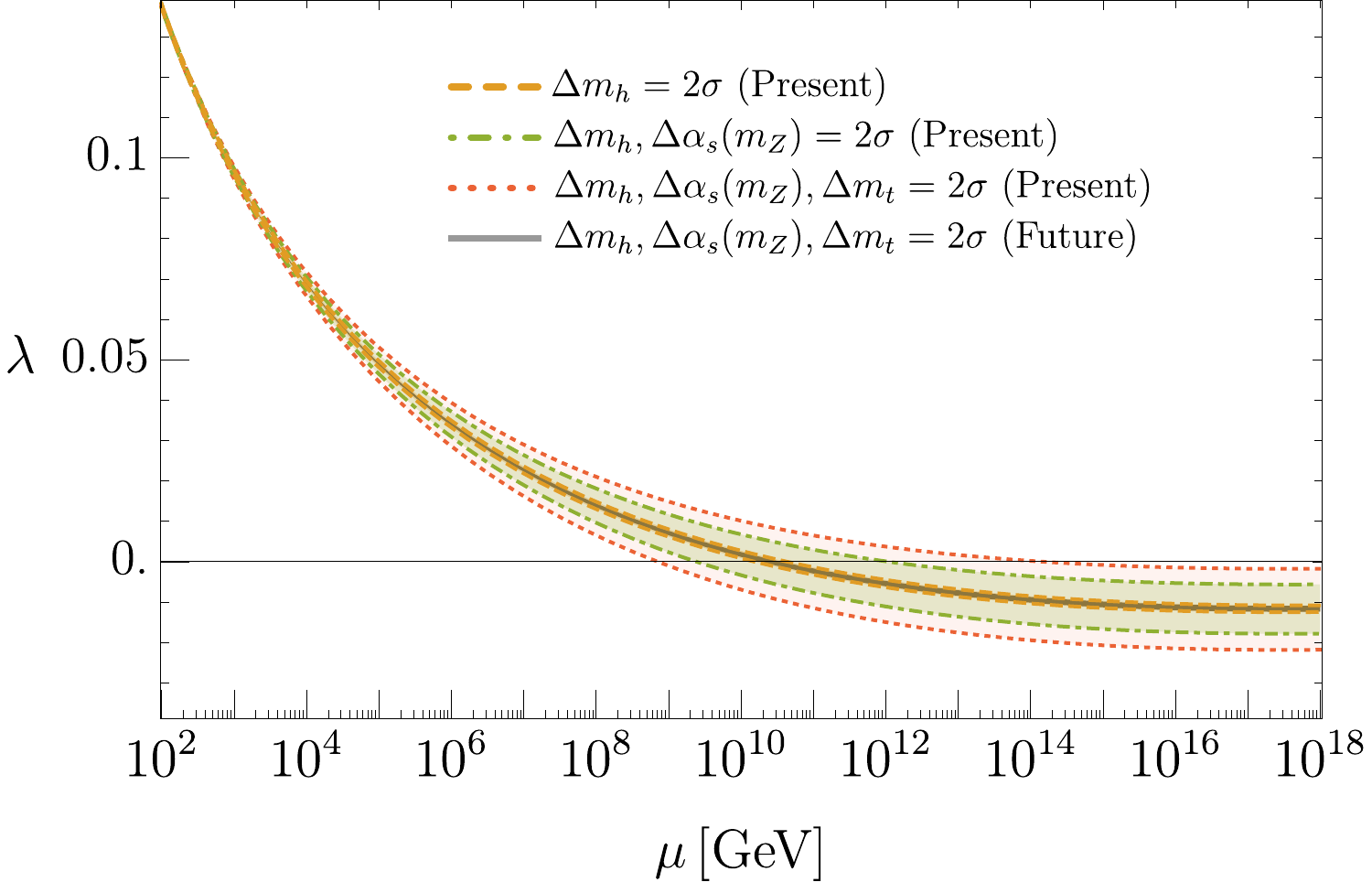} 
    \end{minipage}\hfill
    \begin{minipage}{0.5\textwidth}
        \centering
        \includegraphics[width=.95\textwidth]{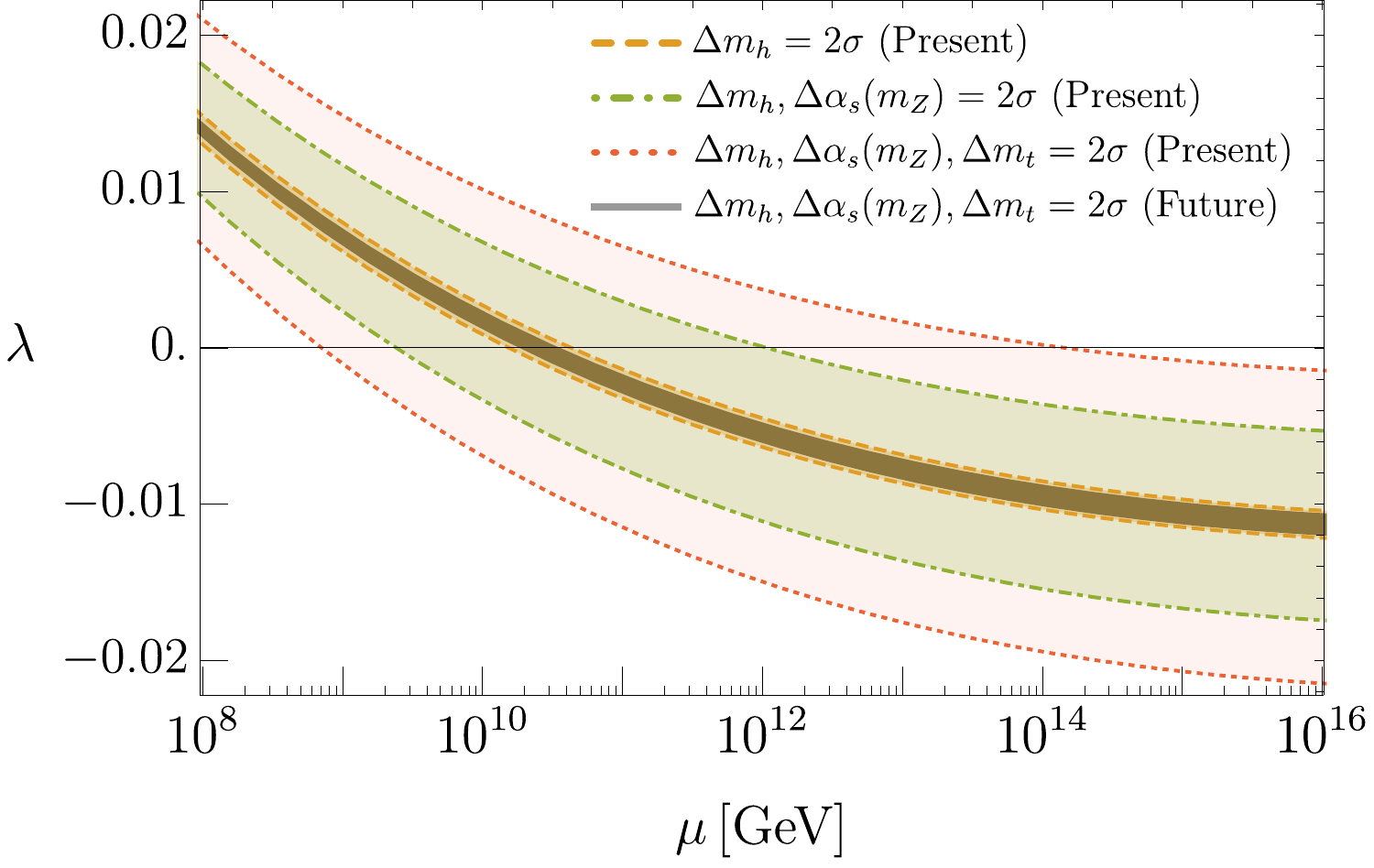} % first figure itself
    \end{minipage}
    \caption{\small  Running of the SM quartic coupling with current and future uncertainties in $m_t$, $\alpha_s(m_Z)$, and $m_h$. Their central values are $m_t = 172.76 ~{\rm GeV}$, $\alpha_s(m_Z) = 0.1179$, and $m_h = 125.10 ~ {\rm GeV}$.}
    \label{fig:λvpPlot}
\end{figure}

The discovery of the Higgs boson at the Large Hadron Collider (LHC) completes the Standard Model (SM). Electroweak symmetry breaking arises from the potential 
\begin{align}
    V_{\rm SM}(H) = -m^2 |H|^2+ \lambda |H|^4,
\end{align}
via the ground state value of the Higgs field $\langle H \rangle =  v \simeq 174 ~\GeV$. The Higgs boson mass is $m_h^2 = 4 \lambda v^2$. 
No other new particles have been discovered at the LHC so far, and in this paper we assume that the SM is valid to very high energies. All the SM couplings can be computed at high energies to high precision, including the Higgs quartic coupling~\cite{Buttazzo:2013uya}. As shown in Fig.~\ref{fig:λvpPlot}, this running indicates that the Higgs quartic coupling vanishes at the scale 
\begin{align}
    \mu_{\lambda} = 10^{9-12} ~{\rm GeV},
    \label{eq:HiggsQuarticScale}
\end{align}
which we call the {\it Higgs quartic scale}. Indeed, within the context of the SM as an effective field theory to very high energies, a key result of the LHC is the discovery of this new mass scale. In this paper we assume that physics beyond the SM first appears at $\mu_{\lambda}$, and the form of the new physics explains why the Higgs quartic is so small at this scale. It is interesting to note that, if dark matter couples to the weak interaction, the recent direct detection experiments have started to explore dark matter masses in the vicinity of the Higgs quartic scale. The mass range to be explored by the next generation of experiments, (\ref{eq:MDMEW}), will probe the entire range of (\ref{eq:HiggsQuarticScale}).

Since the discovery of a Higgs with mass of 125 GeV, several proposals have been made for physics at $\mu_\lambda$ that explains the small quartic coupling, including supersymmetry \cite{Hall:2013eko,Hall:2014vga,Fox:2014moa}, extra dimension~\cite{Gogoladze:2007qm},  Peccei-Quinn symmetry~\cite{Redi:2012ad}, and Higgs Parity symmetry~\cite{Hall:2018let,Dunsky:2019api,Hall:2019qwx,Dunsky:2019upk,Dunsky:2020dhn}.  In this paper we pursue the case of Intermediate Scale Supersymmetry (ISS), where the superpartner mass scale $\tilde{m}$ is of order the Higgs quartic scale. The identification of $\mu_{\lambda}$ with $\tilde{m}$ is natural \cite{Hall:2013eko,Hall:2014vga} since supersymmetry predicts a very small SM Higgs quartic at the scale $\tilde{m}$ for a wide range of supersymmetry breaking parameters. Unlike in \cite{Hall:2013eko,Hall:2014vga}, we study the case of Higgsino or sneutrino Lightest Supersymmetric Particle (LSP) dark matter with mass of order $\tilde{m}$, since this gives a direct detection signal that is correlated with the Higgs quartic scale.

In this paper, we examine the correlation in ISS between the dark matter detection signal via $Z$ exchange and the precision measurement of 
the top quark mass, $m_t$, the strong coupling constant, $\alpha_s(m_Z)$, (and to a lesser extent, of the Higgs boson mass, $m_h$). 
%We that determines $\mu_{\lambda}$.  
A dark matter signal will determine the mass of the LSP and precision measurements will greatly reduce the uncertainties in the Higgs quartic scale. In particular, we find that the discovery of a direct detection signal implies an upper bound on the top quark mass and a lower bound on the strong coupling constant. The effects on the running of the Higgs quartic in reducing the uncertainties in $m_t, \alpha_s(m_Z)$ and $m_h$ are shown by the colored bands in Fig.~\ref{fig:λvpPlot}. Future uncertainties in $m_t ~\, (0.01 {\rm GeV})$, $\alpha_s(m_Z) ~ (0.0001)$, and $m_h ~ (0.01 \, {\rm GeV)}$ from measurements at future lepton colliders \cite{Seidel:2013sqa,Horiguchi:2013wra,Kiyo:2015ooa,Beneke:2015kwa,Gomez-Ceballos:2013zzn}, improved lattice calculations~\cite{Lepage:2014fla}, and the high-luminosity LHC \cite{Cepeda:2019klc}, will substantially reduce the uncertainty in $\mu_{\lambda}$ to within a few tens of percents, as shown by the solid black strip in Fig.~\ref{fig:λvpPlot} which is centered at the current central values of $m_{\rm t}$, $\alpha_s(m_Z)$, and $m_h$.

In 1977, Lee and Weinberg showed dark matter, if coupled to the weak interaction, could be produced in the early universe by freezing-out, losing thermal equilibrium while non-relativistic \cite{Lee:1977ua}. Indeed, they discovered that a heavy neutrino, with a GeV-scale mass, could yield the observed dark matter abundance. Many other electroweak dark matter candidates arising from freeze-out were studied, with masses up to several TeV.  Apparently our proposal of Higgsino or sneutrino dark matter with a mass of  $10^9 - 10^{12}$ GeV leads to a huge overproduction of dark matter. However, we find that the observed abundance can result from freeze-out or freeze-in during a matter-dominated era after inflation.
The inflaton mass must be below the dark matter mass, otherwise the $O(1)$ branching fraction of the inflaton into sparticles leads to an overproduction of dark matter. Then during freeze-out or freeze-in, the inflaton is dissipated by scattering reactions rather than by decays. If the products of the scattering reactions are thermalized at a high enough temperature, freeze-out occurs; otherwise, the abundance is set by freeze-in from non-thermal radiation. Either way, determining the dark matter mass from direct detection will provide a correlation between the reheat temperature after inflation and the inflaton mass.

In section \ref{sec:treeboundarycondition}, building on \cite{Hall:2013eko,Hall:2014vga}, we show that if the UV completion of the SM EFT is provided by ISS,
%with $\tilde{m} \sim \mu_\lambda$,
there is a large region of parameter space where the SM quartic coupling is predicted to be very small at $\tilde{m}$, and hence $\tilde{m} \sim \mu_\lambda$. In section \ref{sec:HiggsinoSneutrinoDM} we compute the present limits on Higgsino and sneutrino dark matter, and compute the reaches expected for XENONnT, LZ, and DARWIN.  We then study the correlation between the dark matter signal and future precision measurements of $m_t$, $\alpha_s(m_Z)$, and $m_h$.  In section \ref{sec:threshold_Corrections} we study how this correlation is affected by supersymmetric threshold corrections to the Higgs quartic coupling in the Minimal Supersymmetric Standard Model (MSSM). We find that these threshold corrections can be significant and derive 
an upper bound on the Higgsino or sneutrino LSP mass as a function of the top quark mass and the strong coupling constant. An observable direct detection signal is predicted for top masses above a critical value.  In section \ref{sec:unifiedsusyBC} we compute the supersymmetric threshold corrections in a scheme where the supersymmetry breaking parameters are constrained to a universal form at unified scales. In section \ref{sec:cosabund} we compute the relic dark matter abundance from freeze-out or freeze-in during a matter dominated era where the inflaton condensate is dissipated by scattering reactions. Finally, we draw conclusions in section \ref{sec:conclusions}.

%\begin{enumerate}
%    \item Lack of supersymmetric particles at the weak scale is not a dead end for SUSY, but in fact, the LHC could have found hints of SUSY via the precise measurements of $m_h$ and $m_t$ via the running of the quartic.
%   \item Higgs quartic appears to vanish at scale $\mu_{\lambda} = 10^{9-12} ~ \GEV$ under the renormalization flow with an uncertainty coming from the top quark mass, higgs mass, and strong coupling constant . Could this be a hint for super symmetry with $\tan \beta \sim 1$ and an intermediate supersymmetry breaking scale $m_{\rm susy}$ near $\mu_{\lambda}$?  Supersymmetry also predicts a vanishing quartic at the scale $m_{\rm susy}$ when $\tan \beta \sim 1$, so it is natural to identify $\mu_{\lambda}$ and $m_{\rm susy}$. 
%    \item  In this paper, we consider such a scenario and note that such a scale of SUSY gives a natural DM candidate in the form of a Higgsino or Sneutrino that could be discovered via direct detection experiments like xenon1t.
%    \item Thus, DM direct detection coupled together with precision measurements at the LHC which pinpoint down the scale $\mu_{\rm lambda}$ and hence $m_{\rm susy}$ could actually discover SUSY at scales much higher than accessible at colliders. 
%   \item Threshold corrections from supersymmetric particles can slightly modify the relationship between $\mu_\lambda$ and $m_{\rm susy}$. Nevertheless, the corrections are generally such that an upper bound on the DM mass and lower bound on $m_{\rm top}$ can be placed. 
%\end{enumerate}  

\section{The Tree-Level Boundary Condition on the SM Quartic Coupling}
\label{sec:treeboundarycondition}
%\section{Higgs quartic coupling at the sparticle mass scale}
%%
%We take the SM to be the effective theory below the Higgs quartic scale, (\ref{eq:HiggsQuarticScale}), which is identified as the scale of supersymmetry breaking, $\tilde{m}$.
We take the SM to be the effective theory below the scale of supersymmetry breaking, $\tilde{m}$.
In this section, we review the tree-level prediction for the SM Higgs quartic coupling, $\lambda_{\rm tree}$. At scale $\tilde{m}$, we assume that there is no gauge symmetry breaking and the theory contains a single pair of Higgs doublets, $(H_u, H_d)$, and no weak singlets or triplets which have a zero hypercharge and couple to the Higgs doublets. For a wide range of parameters of this Higgs sector, we find $\lambda(\tilde{m}) \ll 0.01$; remarkably there are large regions with $\lambda(\tilde{m}) \lesssim 0.001$, and the supersymmetry breaking scale $\tilde{m}$ may be identified with the Higgs quartic scale $\mu_\lambda$.
%We then consider to what extent threshold corrections from sparticles near the scale $\tilde{m}$ modify the tree-level result. Finally, we review how the renormalization running of $\lambda$ from low to high scales in the Standard Model (SM) EFT also leads to $\lambda \ll 1$ at a scale $\mu_{\lambda} = 10^{9-12} ~ \GEV$, depending predominantly on the value of the top quark mass. We discuss the natural conclusion of identifying $\mu_\lambda$ with $\tilde{m}$.

%We take the EFT below the scale $\tilde{m}$ to be the SM, and above the scale $\tilde{m}$ to be the MSSM. 
The Higgs potential is
\begin{align}
    \label{eq:MSSM_potential}
    V(H_u,H_d) =& \, (\mu^2 + m_{H_u}^2)H_u^\dagger H_u + (\mu^2 + m_{H_d}^2)H_d^\dagger H_d + (B\mu \; H_u H_d + {\rm h.c.}) \nonumber \\
    &+ \frac{g^2}{8} (H_u^\dagger \vec{\sigma} H_u + H_d^\dagger \vec{\sigma} H_d)^2 + \frac{g'^2}{8}(H_u^\dagger  H_u - H_d^\dagger  H_d)^2,
\end{align}
where $\mu$ is the supersymmetric Higgs mass parameter, while $m_{H_u}^2,  m_{H_d}^2$, and $B\mu$ are supersymmetry-violating mass parameters. 
These parameters are all taken real, without loss of generality, and have sizes determined by the scale of supersymmetry breaking, $\tilde{m}$. The constants $g$ and $g'$ are the $SU(2)$ and $U(1)$ gauge couplings. Requiring electroweak symmetry to be unbroken at $\tilde{m}$ and one combination of the Higgs doublets to be much lighter than $\tilde{m}$ requires that $\mu^2 + m_{H_{u,d}}^2$ are both positive. The fine tune for a light doublet requires that $B \mu$ is taken to be the geometric mean of $\mu^2 + m_{H_{u,d}}^2$. The light SM Higgs doublet is
\begin{align}
    \label{eq:H}
    H = \sin \beta \, H_u + \cos \beta \, H_d^\dagger,
\end{align}
where
$\tan^2 \beta = (\mu^2 + m_{H_{d}}^2)/(\mu^2 + m_{H_{u}}^2)$, and we take $\beta$ in the first quadrant.

\begin{figure}[tb]
    \begin{center}
        \includegraphics[width=0.6\textwidth]{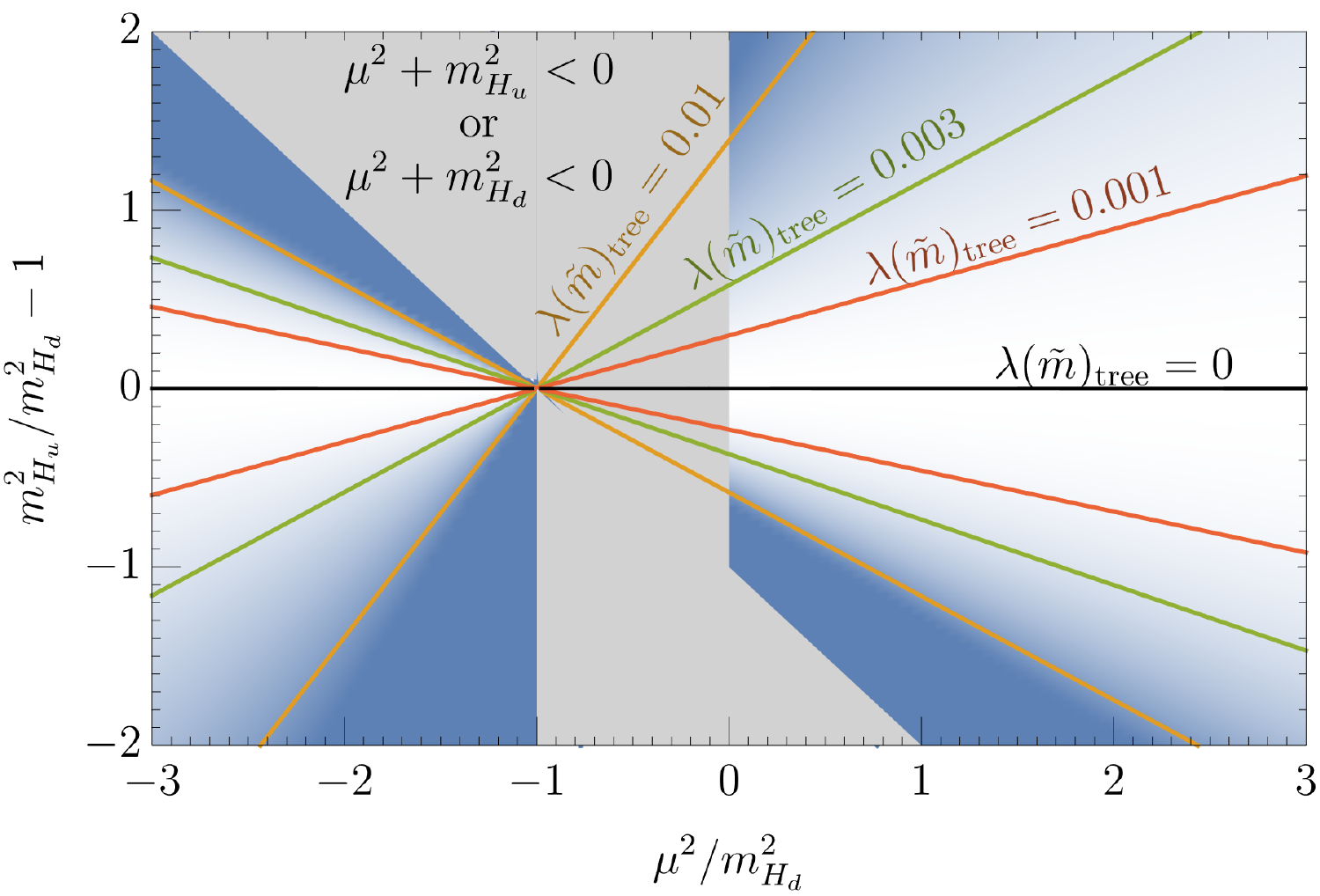}
    \end{center}
    \caption{Regions of parameter space showing the smallness of the ISS tree-level prediction for the Higgs quartic coupling at the scale $\tilde{m}$. $\lambda(\tilde{m})_{\rm tree}$ is less than  $10^{-3}$ if $\mu$ is much greater than $m_{H_u}$ and $m_{H_d}$, or if $m_{H_u}$ and $m_{H_d}$ are nearly degenerate. The tree-level prediction is zero when $m_{H_d}^2 = m_{H_u}^2$, as indicated by the black horizontal line. In the {\color{gray} \textbf{gray}} region, one of the Higgs doublets has a negative mass squared. With Higgsino or sneutrino LSP, the {\color{c1} \textbf{blue}} region is excluded by XENON1T.}
    \label{fig:λISSPlot}    
\end{figure}

Matching the two theories at $\tilde{m}$ gives the tree-level value for $\lambda(\tilde{m})$
\begin{align}
    \label{eq:treeQuartic}
    \lambda(\tilde{m})_{\rm tree} = \frac{g_2(\tilde m)^2 + g'(\tilde m)^2}{8} \cos^2 2\beta
\end{align}
with
%Furthermore, assuming that $\mu, m_{H_u}, m_{H_d}$ and $B$ are much larger than the electroweak scale implies the determinant of the $(H_u ~ H_d)$ mass matrix nearly vanishes in order that it contain a nearly massless eigenvalue, namely, the SM Higgs mass. It follows that $\mu B$ is the geometric mean of the first two coefficients in Eq. \eqref{eq:MSSM_potential}, allowing the mixing angle between $H_d$ and $H_u$ to be written as
%
\begin{align}
    \label{eq:cossq2beta}
    \cos^2 2\beta = \left(\frac{m_{H_u}^2 - m_{H_d}^2}{m_{H_u}^2 + m_{H_d}^2 + 2\mu^2}\right)^2.
\end{align}
% 
%With $\tilde{m}$ of order $\mu_\lambda$, 
ISS gives $0 \leq\lambda(\tilde{m})_{\rm tree} \leq (g_2(\tilde m)^2 + g'(\tilde m)^2)/8 \simeq 0.06 $ and hence at tree level $\tilde{m} \lesssim \mu_\lambda$.  Furthermore, over a wide range of values for $m_{H_u}^2,m_{H_d}^2$, and $\mu$ the $\cos^2 2 \beta$ factor gives a significant further suppression of $\lambda(\tilde{m})_{\rm tree}$, as shown in Fig.~\ref{fig:λISSPlot}. Indeed, $\cos 2\beta \rightarrow 0$ in the limit that either $\mu^2 \gg |m_{H_{u,d}}^2|$ or $m_{H_u}^2 \rightarrow m_{H_d}^2$; in these limits $\tilde{m}$ is identified with $\mu_\lambda$.
The gray-shaded region is excluded since $\mu^2 + m_{H_u}^2<0$ or $\mu^2 + m_{H_d}^2<0$ and there is no stable vacuum with a large hierarchy between the weak scale and the supersymmetry breaking scale.
%In the vertical mauve band the condition $\mu^2 + m_{H_d}>0$ is inconsistent with the physical condition that $\mu^2 >0$, and hence is also excluded.
In the blue-shaded region, where $\lambda(\tilde{m})_{\rm tree}>0.01$, $\tilde{m}$ is predicted to be below a few $10^9$ GeV. As we will see in the next section, the Higgsino or sneutrino LSP then gives too large a direct detection rate.
However, there is a remarkably large region of parameter space in Fig.~\ref{fig:λISSPlot} with $\lambda(\tilde{m})_{\rm tree} < 0.003$.

%\subsection{Renormalization group running and standard model parameters}
%%

\section{Direct Detection of Dark Matter}
\label{sec:HiggsinoSneutrinoDM}

%this paper we analyze the supersymmetric threshold corrections to the Higgs quartic coupling in ISS, deriving upper bounds on the expected Higgsino or sneutrino LSP mass relative to the Higgs quartic scale. An observable direct detection signal is predicted for top masses below a critical value.

In this section, we discuss direct detection of the Higgsino or sneutrino LSP dark matter in nuclear recoil experiments and show that detection rates are correlated with SM parameters through the connection between $\tilde{m}$ and $\mu_{\lambda}$. An observable direct detection signal is predicted for top masses below a critical value.

\subsection{Higgsino or sneutrino dark matter}
\subsubsection{Higgsino dark matter}
%The Higgsino is the LSP when $\mu$ is the smallest SUSY mass parameter.
The neutral components and the charged component of the Higgsino are degenerate in mass in the electroweak symmetric limit. With elecroweak symmetry breaking, the charged component becomes heavier than the neutral components by $O(100)$ MeV via one-loop quantum corrections~\cite{Cirelli:2005uq}. The neutral components slightly mix with the bino and the wino and obtain a small mass splitting
\begin{align}
    \Delta m \sim \frac{g^2 v^2}{M_2} \approx 10 \, {\rm keV} \left(\frac{M_2}{10^9 \, {\rm GeV}}\right)^{-1}.
\end{align}
The two mass eigenstates are Majorana fermions. For a soft mass scale above $\sim 10^{9}$ GeV, however, the splitting is smaller than the typical nucleon recoil energy of $O(10-100)$ keV, and the Majorana nature does not affect the rate of dark matter signals. Specifically, $Z$ boson exchange leads to the up-scattering of the ligher state into the heavier state, which almost behaves as scattering of a Dirac fermion.

\subsubsection{Sneutrino dark matter}
The sneutrino is lighter than its charged $SU(2)$ partner because of electroweak symmetry breaking and quantum corrections. The two components of the sneutrino obtain a small mass splitting from the $A$ term of the Majorana neutrino mass term,
\begin{align}
    \Delta m \sim \frac{A m_\nu}{m_{\tilde{\nu}}},
\end{align}
which is negligibly small. Sneutrino dark matter interacts with nucleon via $Z$ boson exchange as a complex scalar field.

If the slepton and squark masses are universal at the unification scale, the sneutrino cannot be the LSP because renormalization running makes the right-handed stau the lightest among them. Non-universality is required for the sneutrino LSP. We note that the sneutrino LSP is consistent with $SU(5)$ unification, since the sneutrinos and the right-handed sleptons are not unified, and the right-handed down type squarks become heavier than the sneutrinos by renormalization running.

\begin{figure}[tb]
    \begin{center}
        \includegraphics[width=0.7\textwidth]{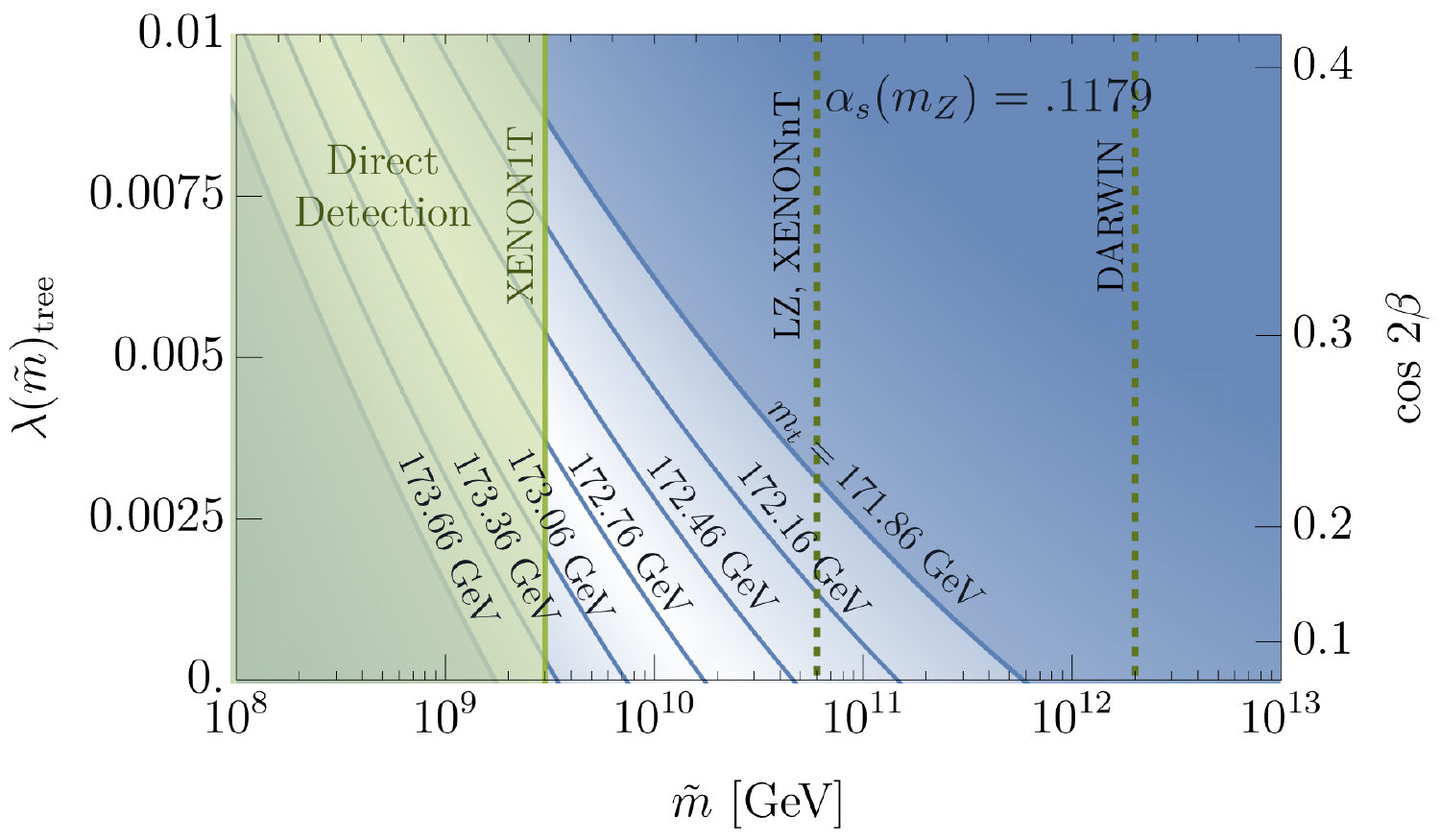}
    \end{center}
    \caption{Prediction for the top quark mass as a function of the sparticle mass scale, $\tilde{m}$, and the tree-level Higgs quartic coupling at $\tilde{m}$. Contours of $m_{\rm top}$ span $3 \sigma$ above and below the current central value for $m_{\rm top}$, $(172.76 \pm 0.30) \, {\rm GeV}$. For Higgsino or sneutrino LSP dark matter, the {\color{c3} \textbf{green}} shaded region is excluded by XENON1T and dotted {\color{c3} \textbf{green}} lines show the sensitivities of future experiments.  Values of $m_t$ are experimentally disfavored in the dark {\color{c1} \textbf{blue}} region}
    \label{fig:λvsmtildeTreePlot}    
\end{figure}

\subsection{Direct detection rate and standard model parameters}

Both Higgsino and sneutrino dark matter scatter with nuclei, with an effective dark matter-nucleon scattering cross section given by
\begin{align}
    \sigma_n = \frac{G_F^2 m_n^2}{2\pi} \left[\frac{(A-Z) - (1-4{\rm sin}^2 \theta_W)Z}{A}\right]^2,
\end{align}
where $G_F$ is the Fermi constant, $m_n$ is the nucleon mass, $A$ is the mass number, $Z$ is the atomic number, and $\theta_W$ is the Weinberg angle.
The current constraint by XENON1T~\cite{Aprile:2018dbl} and the future sensitivities of LZ with an exposure of 15 ton$\cdot$year, XENONnT with an exposure of 20 ton$\cdot$year, and DARWIN with an exposure of 1000 ton$\cdot$year~\cite{Akerib:2018lyp,Aprile:2015uzo,Aalbers:2016jon} are given by
\begin{align}
    \sigma_n &< 2\times 10^{-11}{\rm GeV}^{-2} \frac{m_{\rm DM}}{10^{10}~{\rm GeV}}~~(\text{XENON1T, current}). \\
    \sigma_n &< 1\times 10^{-12}{\rm GeV}^{-2} \frac{m_{\rm DM}}{10^{10}~{\rm GeV}}~~(\text{LZ, XENONnT, future}). \\
    \sigma_n &< 4\times 10^{-14}{\rm GeV}^{-2} \frac{m_{\rm DM}}{10^{10}~{\rm GeV}}~~(\text{DARWIN, future}) ,
\end{align}
which translates into the constraint on and the sensitivity to the Higgsino or sneutrino dark matter mass of
\begin{align}
    m_{\rm DM} &> 3\times 10^{9}~{\rm GeV}~~(\text{XENON1T, current}), \\
    m_{\rm DM} &> 6\times 10^{10}~{\rm GeV}~~(\text{LZ, XENONnT future}), \\
    m_{\rm DM} &> 2\times 10^{12}~{\rm GeV}~~(\text{DARWIN, future}).
\end{align}

Once dark matter signals are found in recoil experiments, within the framework of Higgsino or sneutrino dark matter in ISS, the dark matter mass is fixed from the observed signal rates.
Since $\lambda(\tilde{m})_{\rm tree}$ is positive and $m_{\rm DM} = m_{\rm LSP}< \tilde{m}$, we obtain a bound on SM parameters including an upper bound on the top quark mass.
Conversely, for given SM parameters,  $m_{\rm DM}$ is bounded from above.
The prediction for the top quark mass for given $\tilde{m}$ and $\lambda(\tilde{m})_{\rm tree}$ is shown in Fig.~\ref{fig:λvsmtildeTreePlot}. The right vertical axis shows ${\rm cos}2\beta$ corresponding to $\lambda(\tilde{m})_{\rm tree}$.
For a given $m_{\rm DM}$, the prediction on $m_t$ for $\lambda(\tilde{m})_{\rm tree}=0$ and $\tilde{m}=0$ can be understood as an upper bound on $m_t$. For a given $m_t$, $\tilde{m}$ such that $\lambda(\tilde{m})_{\rm tree}=0$  in an upper bound on $m_{\rm DM}$.
To obtain those bounds precisely, we include threshold corrections to $\lambda(\tilde{m})$ in the next section.

\section{Including Threshold Corrections to the Higgs Quartic}
\label{sec:threshold_Corrections}

The full prediction for $\lambda(\tilde{m})$ in ISS is
\begin{align}
    \lambda(\tilde{m}) = \lambda(\tilde{m})_{\rm tree} + \delta \lambda(\tilde{m}),
\end{align}
where $\lambda_{\rm tree}$ is the the tree-level result, \eqref{eq:treeQuartic}, and $\delta \lambda$ the quantum corrections that arise on integrating out heavy sparticles.  The largest contributions arise from sparticles with the largest couplings to the light Higgs; hence the most important mass parameters are the masses of the third generation doublet squark $m_{\tilde{q}}$, the third generation up-type squark $m_{\tilde{\bar{u}}}$,
the bino $M_1$, the wino $M_2$, the heavy Higgs $m_A$, and the $A$ term of the top quark yukawa $A_t$.

We choose the matching scale to be the lighter of $m_{\tilde{q}}$ and $m_{\tilde{\bar{u}}}$, which we denote as $m_-$. Since quantum corrections are greater than $\lambda_{\rm tree}$ only for ${\rm tan}\, \beta \simeq 1$, we neglect corrections which vanish in this limit.
Using the results in~\cite{Giudice:2011cg}, the corrections are given by
\begin{equation}
\label{eq:deltalambda}
\begin{split}
     32 \pi^2\delta \lambda(m_-) &=  3 y_t^4 \left( \ln \frac{m_{\tilde{q}}^2}{m_- ^2} + \ln \frac{m_{\tilde{\bar{u}}}^2}{m_- ^2} + 2 X_t F\left(\frac{m_{\tilde{q}}}{m_{\tilde{\bar{u}}}}\right) - \frac{X_t^2}{6} G\left(\frac{m_{\tilde{q}}}{m_{\tilde{\bar{u}}}}\right) \right)  \\
     &-\frac{1}{4}\left(g^{'4} + 2g^{'2} g^2 + \frac{16}{3} g^4 \right)  - \frac{4}{3} g^{'4} f_1\left(\frac{M_1}{\mu}\right) - 4g^{4}f_1\left(\frac{M_2}{\mu}\right)  - \frac{8}{3}g^{'2}g^{2} f_2\left(\frac{M_1}{\mu},\frac{M_2}{\mu}\right)   \\
      &-(g^{'4} + 2 g^{'2}g^2+ 3 g^4) \ln\left(\frac{\mu}{m_-}\right)  + \frac{1}{8}\left(g^{'4} + 2 g^{'2} g^2 + 3 g^{4} \right)\ln \frac{m_A^2}{m_-^2}.
\end{split}
\end{equation}
Here, $X_t \equiv (A_t-\mu)^2/m_{\tilde{\bar{u}}} m_{\tilde{\bar{q}}}$, and the functions $F, G, f_1,f_2$ are given by
\begin{align}
    F(x) &= \frac{2x \ln x}{x^2 - 1},~~
    G(x) = \frac{12x^2(1-x^2 + (1+x^2)\ln x)}{(x^2-1)^3}, \nonumber \\ 
    f_1(x)&= \frac{3(x+1)^2}{8 (x-1)^2} + \frac{3(x-3) x^2 {\rm ln}x}{4(x-1)^3}, \nonumber \\
    f_2(x,y)& = \frac{3(1 + x + y - xy)}{8(x-1) (y-1)} + \frac{3 x^3 {\rm ln} x}{4(x-1)^2(x-y)} -\frac{3 y^3 {\rm ln} y}{4(y-1)^2(x-y)}.
\end{align}
They are normalized so that they are unity when the arguments are unity. For a degenerate mass spectrum and negligible $X_t$, $\delta\lambda(m_-)\simeq - 0.002$.

\begin{figure}[tb]
    \centering
    \begin{minipage}{0.5\textwidth}
        \centering
        \includegraphics[width=.95\textwidth]{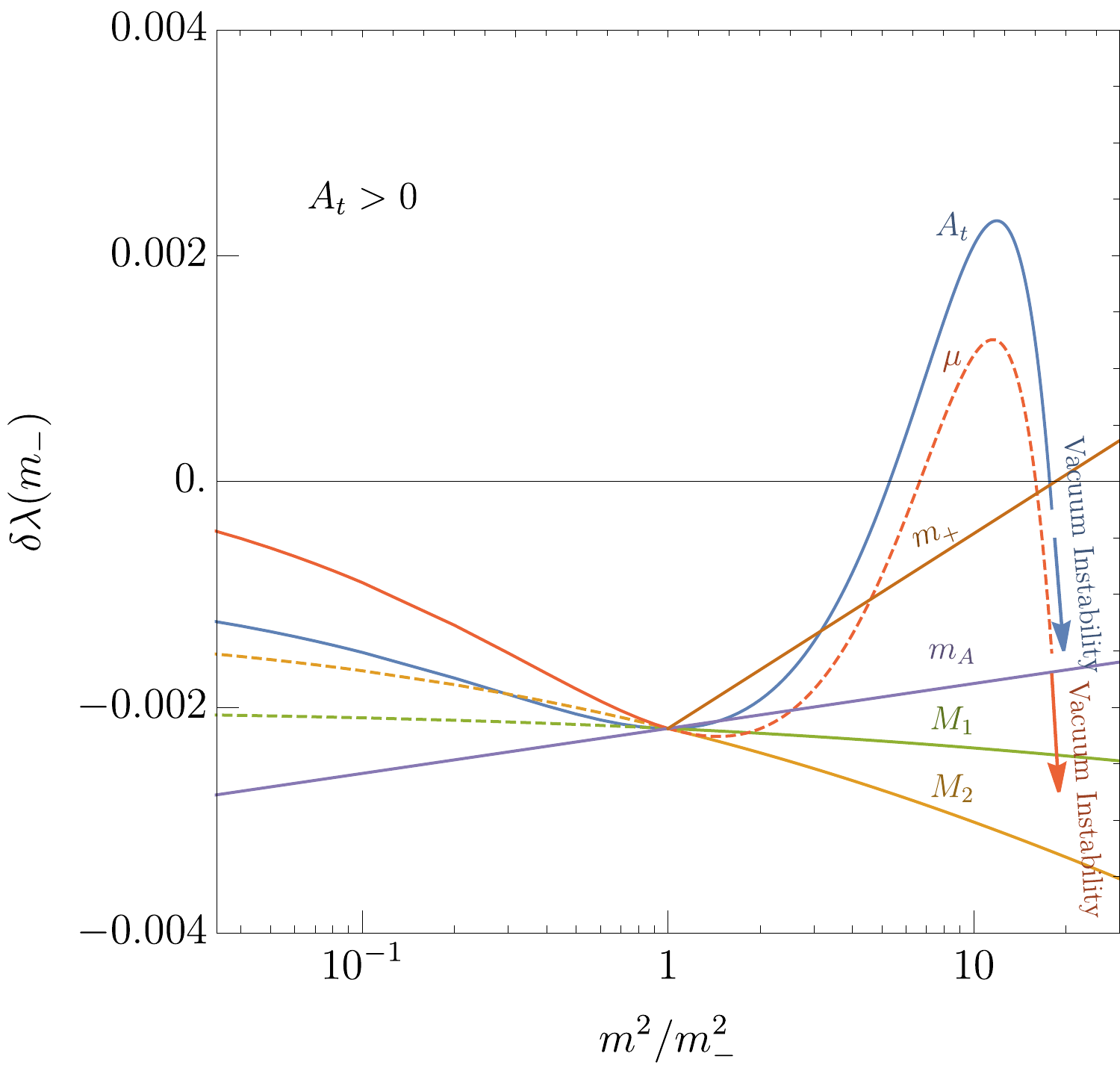} 
    \end{minipage}\hfill
    \begin{minipage}{0.5\textwidth}
        \centering
        \includegraphics[width=.95\textwidth]{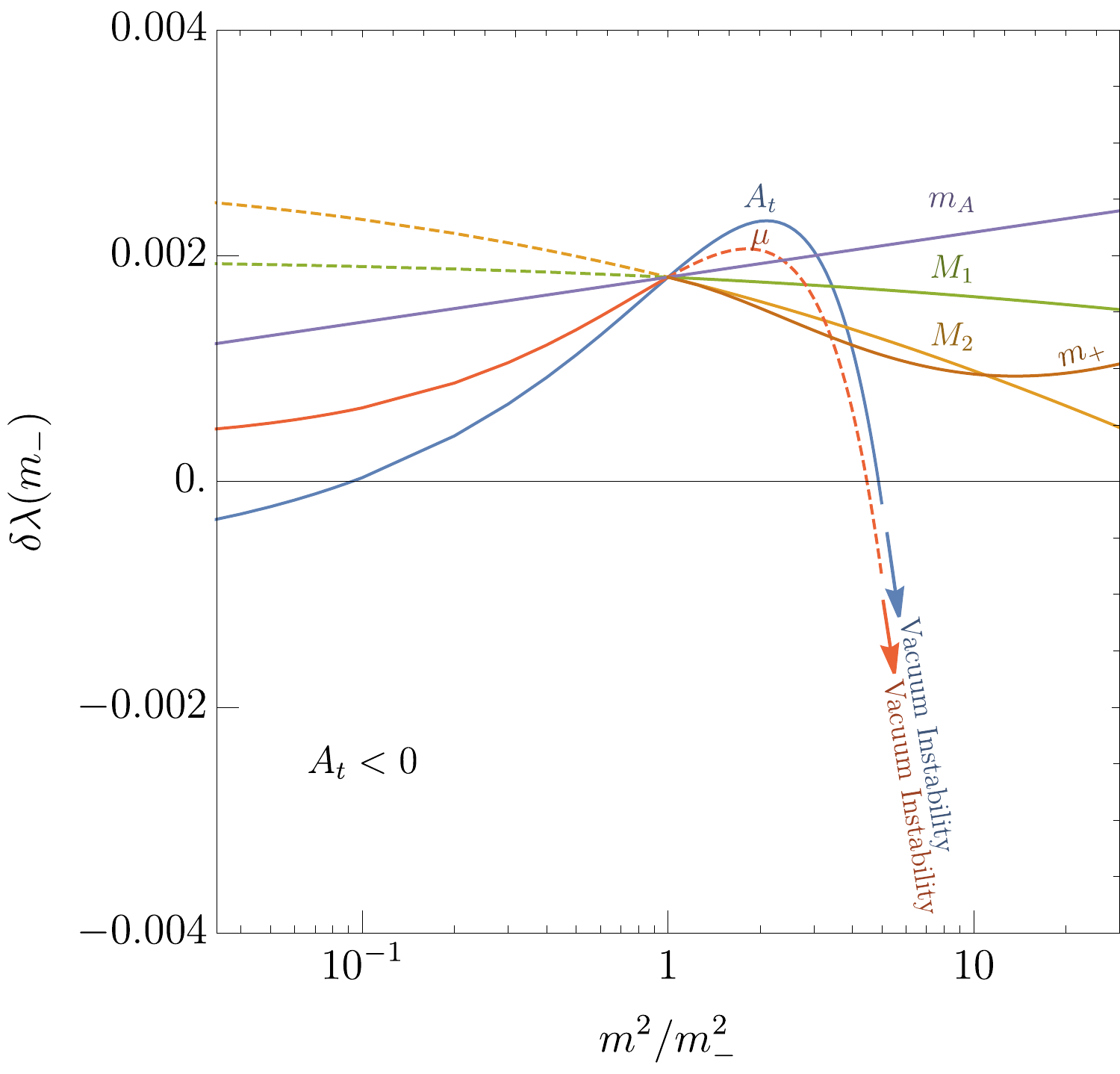} % first figure itself
    \end{minipage}
    \caption{\small Threshold corrections to the Higgs quartic coupling as a function of sparticle mass parameters. The six curves correspond to $m= (A_t, \mu, m_+, m_A, M_1, M_2)$ with the remaining five parameters fixed at $m_- = {\rm{min}} (m_{\tilde{q}}, m_{\tilde{\bar{u}}})$. The Higgsino can be the LSP on the solid curves, but is not the LSP on the dashed part of the curves for $\mu, M_1$ and $M_2$.
    %If Higgsino is the LSP, then $\mu/m_- \leq 1$, as signified by the solid shading for the $\mu$ curve. If Sneutrino is the LSP, then $\mu/m_-$ can be greater than $1$, as signified by the dashed shading for the $\mu$ curve. Likewise, if $M_1/m_-$ or $M_2/m_-$ are less than $1$, then the Higgsino cannot be the LSP, as signified by the dashed shading. 
    \textbf{Left} $A_t > 0$. Vacuum instability occurs when $A_t, \mu \gtrsim 4.2 m_-$.  \textbf{Right} $A_t < 0$. Vacuum instability occurs when $|A_t|, \mu \gtrsim 2.2 m_-$.}
    \label{fig:lambdaQC}
\end{figure}

In Fig.~\ref{fig:lambdaQC}, we evaluate Eq.~\eqref{eq:deltalambda} and show how $\delta \lambda$ varies as a function of sparticle masses.  The left and right panels correspond to $A_t$ positive and negative, respectively. Each curve corresponds to varying one of $(A_t, \mu, m_+, m_A, M_1, M_2)$, while keeping all the others fixed at $m_-$. With all these parameters near $m_-$, the correction is $\delta \lambda(m_-)\simeq -0.002$ for $A_t >0$ or $+0.002$ for $A_t <0$. For $|X_t| \gtrsim 10 m_-$, the electroweak vacuum is unstable, as shown by the sudden discontinuation of the $A_t$ and $\mu$ curves. The bound on $X_t$ from the instability is derived in Appendix~\ref{app:instability}. 
%If the Higgsino is the dark matter, then the Higgsino is the LSP and hence the $\mu$ curve exists only for $\mu < m_-$, as indicated by the solid red curve. If the sneutrino is the dark matter, then the Higgsino is not the LSP and hence $\mu$ can be greater than $m_-$, as shown by the dashed red curve.
The Higgsino can be the LSP on the solid curves, but is not the LSP on the dashed part of the curves for $\mu, M_1$ and $M_2$. The slepton mass parameter $m_{\tilde{l}}$ may be taken small enough to give sneutrino LSP anywhere on the lines.

%%%%%
%%%%%
%%%%%
%%%%%

%%%%%
%%%%%
\begin{figure}[tb]
    \begin{center}
          \includegraphics[width=0.65\textwidth]{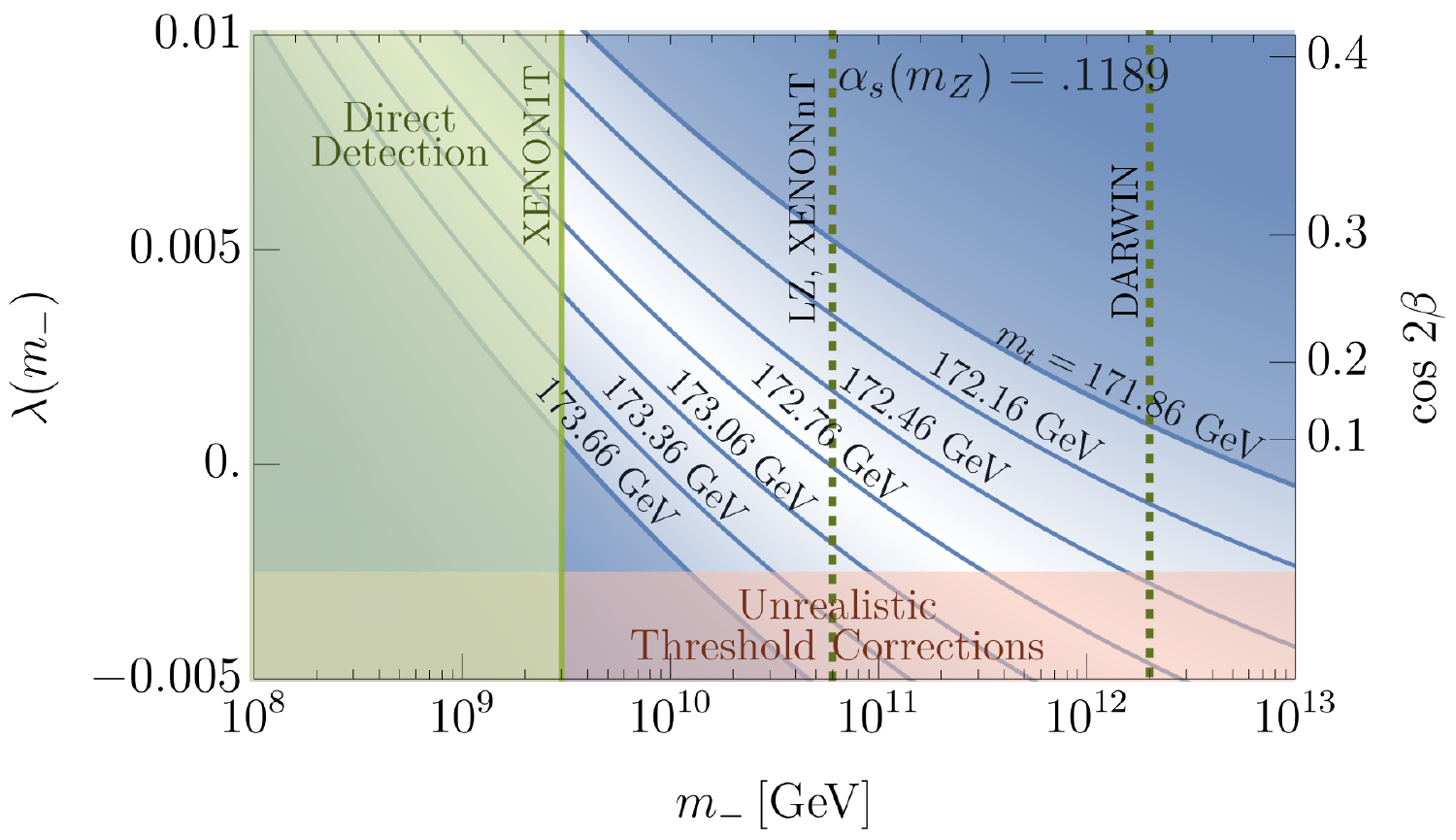}
        \includegraphics[width=0.65\textwidth]{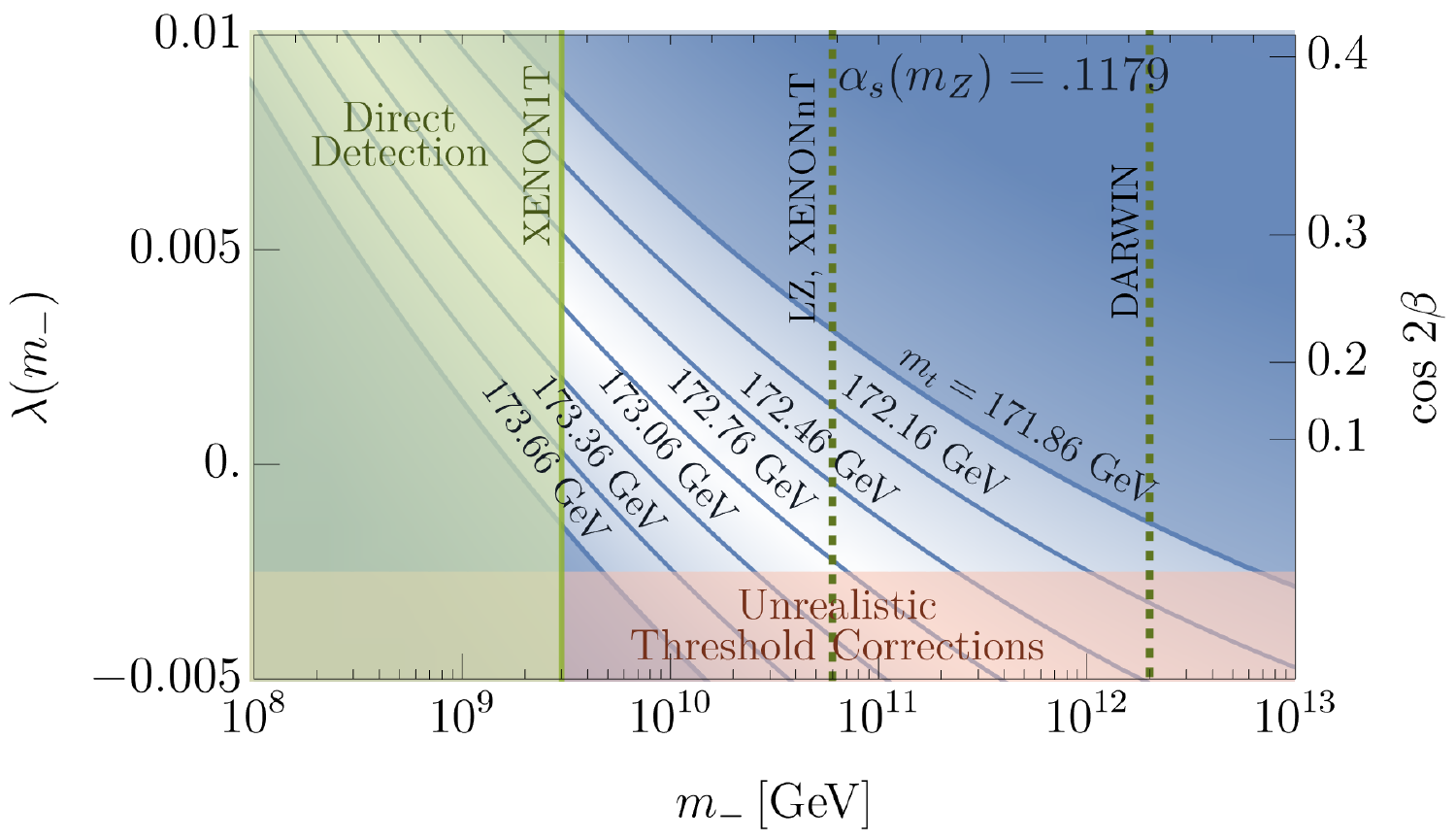}
        \includegraphics[width=0.65\textwidth]{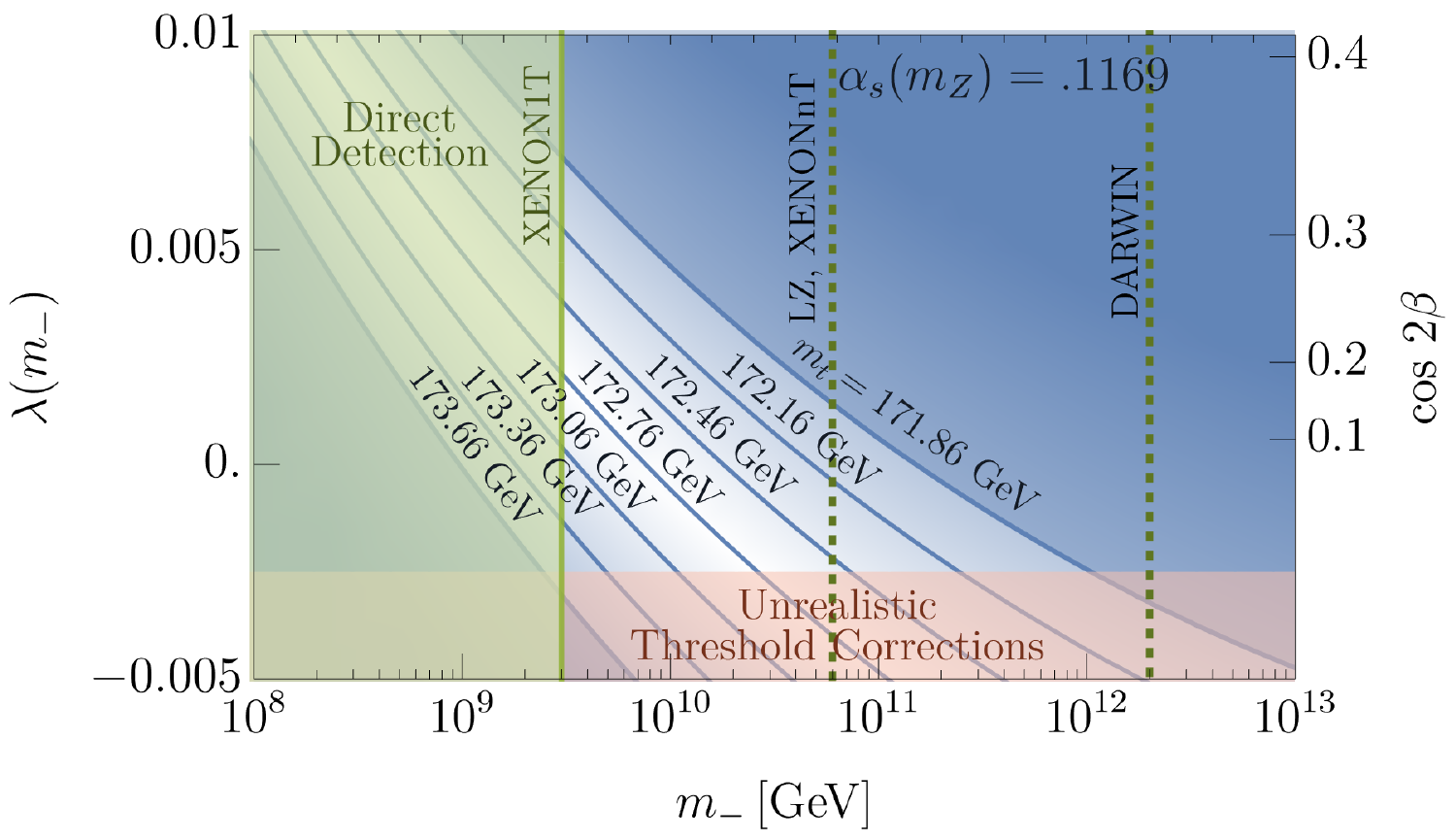}
    \end{center}
     \caption{Prediction for the top quark mass as a function of $m_- = \rm{min} (m_{\tilde{q}}, m_{\tilde{\bar{u}}})$ and the Higgs quartic coupling at $m_-$. Contours of $m_t$ span $3 \sigma$ above and below the current central value for $m_t$, $(172.76 \pm 0.30) \, {\rm GeV}$. The {\color{c4} \textbf{red}} shaded region requires unrealistically large negative supersymmetric threshold corrections to the quartic coupling. The {\color{c3} \textbf{green}} shaded region and the green dotted lines are as in Fig.~\ref{fig:λvsmtildeTreePlot}. Values of $m_t$ are experimentally disfavored in the dark {\color{c1} \textbf{blue}} region.}
    \label{fig:λvsmtildeLoopPlot}    
\end{figure}
%%%%%
%%%%%

%\textit{improve discussion on how we get the upper bound on top quark mass}

% Depending on the spectra of sparticles near $\tilde{m}$, $\delta \lambda$ can be relatively large compared to $\lambda_{\rm tree}$, as shown in Fig. . Here we use the 3-loop quantum corrections calculated in \cite{}, and scan over

%We can obtain an upper bound on the top quark mass in the following way. Since the DM mass is always smaller than $m_-$, the signal rates give a lower bound on $m_-$. Assuming no hierarchy among the important mass parameters listed above, the value of $\lambda(m_-)$ is $-0.002$ plus a tree-level positive contribution from ${\rm tan}\beta \neq 1$, and hence $\lambda(m_-)\gtrsim -0.002$. Together with the lower bound on $m_-$, we obtain an upper bound on the top quark mass.

We show contours of the prediction for $m_t$ in the $(m_-, \lambda(m_-))$ plane in Fig.~\ref{fig:λvsmtildeLoopPlot}, with the strong coupling constant varied within $\pm 1\sigma$ uncertainty from its central value in the top and bottom panels. The right axis shows ${\rm cos}2\beta$ corresponding to $\lambda(m_-)$ when $\delta\lambda\ll \lambda_{\rm tree}$. The lower bound on the dark matter mass from XENON1T is shown in green, and the lower bound on threshold corrections to $\lambda(m_-)$ is shown in red. Together, these bounds require $m_t \lesssim 174.2 ~{\rm GeV}$. The reach of the DARWIN experiment, shown by the dashed green line, will strongly limit the top quark mass to $m_t \lesssim 172.4 \, {\rm GeV}$, if no signals are found.
For the central values of SM parameters, the dark matter mass is required to be below $7\times 10^{10}$ GeV, and LZ and XENONnT can cover most of the parameter space.

%\textit{What do you think if we just put $\cos 2\beta$ on the right axis of Fig. 5, and say that if $\delta \lambda \ll \lambda_{\rm tree}$, then the value of $\cos 2\beta$ on the right axis gives the corresponding value of $\lambda$?}

% %%
% \begin{figure}[h]
%     \begin{center}
%         \includegraphics[width=0.6\textwidth]{Figures/thresholdsVsMtildePlot.pdf}
%     \end{center}
%     \caption{}
%     \label{fig:thresholdsVsMtilde}    
% \end{figure}
% %%

\begin{figure}[tb]
    \centering
    %\begin{minipage}{0.5\textwidth}
        %\centering
        \includegraphics[width=.5\textwidth]{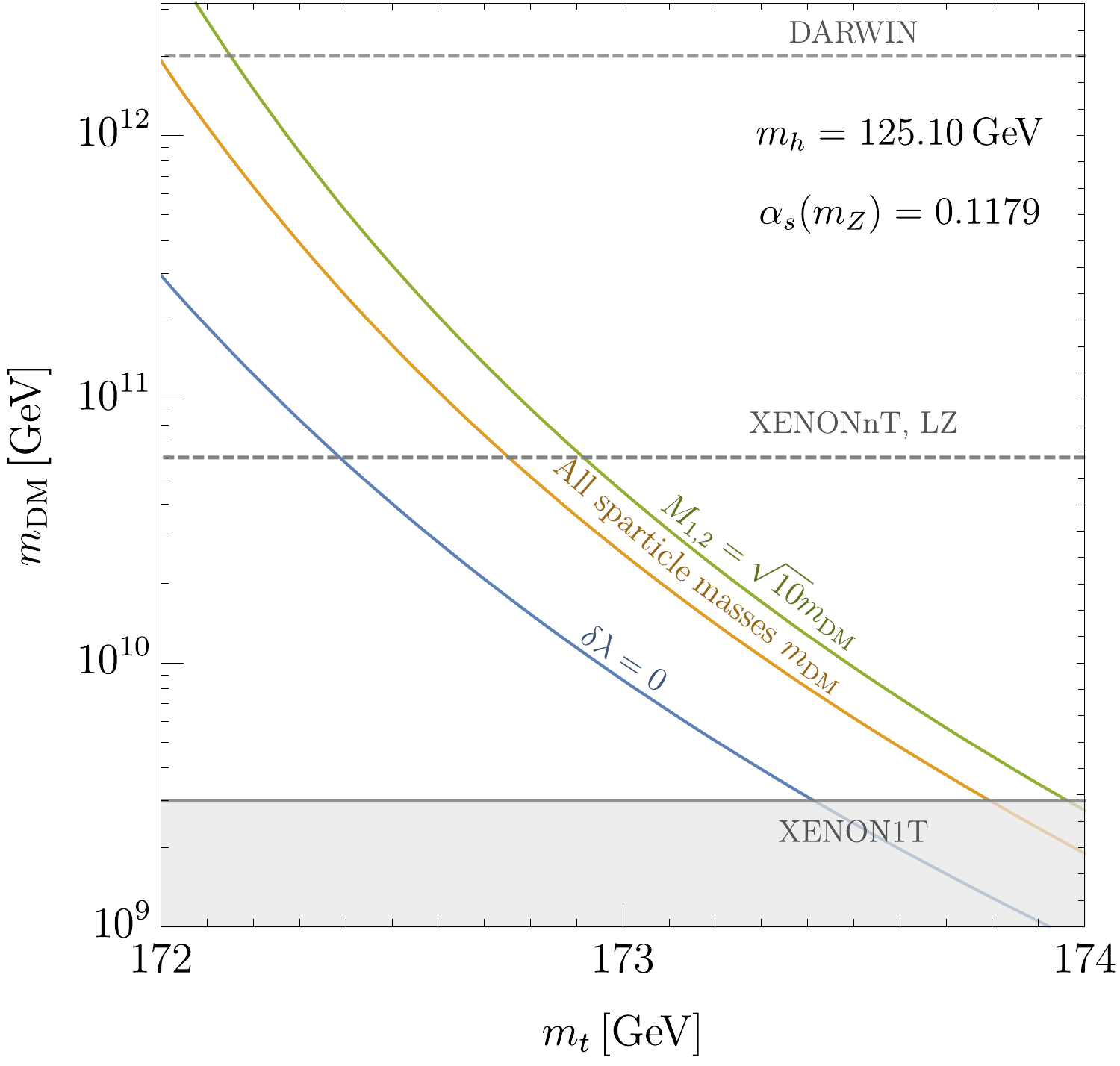} 
    %\end{minipage}\hfill
    \caption{\small Upper bound on the dark matter mass $m_{\rm DM}$ as a function of the top quark mass $m_{t}$ for a range of typical threshold corrections. The {\color{c1} \textbf{blue}} curve shows the bound when the threshold corrections are zero, the {\color{c2} \textbf{orange}} curve when the sparticle spectra are degenerate $m_-$, and in {\color{c3} \textbf{green}}, when $M_{1,2} = \sqrt{10}m_-$. Equivalently, the figure shows an upper bound on $m_t$ as a function of $m_{\rm DM}$.}
    \label{fig:mTopvsmDM}
\end{figure}

The bounds on the dark matter and top quark masses may be relaxed by hierarchical sparticle masses. As shown in Fig.~\ref{fig:lambdaQC}, large wino or bino masses give negative threshold corrections to the quartic coupling, thereby relaxing the upper bounds on the top quark mass and the dark matter mass. In Fig.~\ref{fig:mTopvsmDM}, we show the upper bound on the dark matter mass as a function of the top quark mass or, equivalently, the upper bound on the top quark mass as a function of the dark matter mass. The blue curve is without threshold corrections, the orange curve has threshold corrections for a degenerate mass spectrum with $A_t \simeq \mu$, and on the green curve, the degeneracy is lifted by taking $M_{1,2} = \sqrt{10} m_-$. With this hierarchy, the upper bound on the dark matter mass is relaxed by a factor of $2$, and that on the top quark mass is relaxed by $100$ MeV. (Assuming a high mediation scale of supersymmetry breaking, a larger hierarchy is destabilized by quantum corrections from the gauginos to the soft scalar masses.)

\begin{figure}[tb]
    \centering
    %\begin{minipage}{0.5\textwidth}
        %\centering
        \includegraphics[width=.5\textwidth]{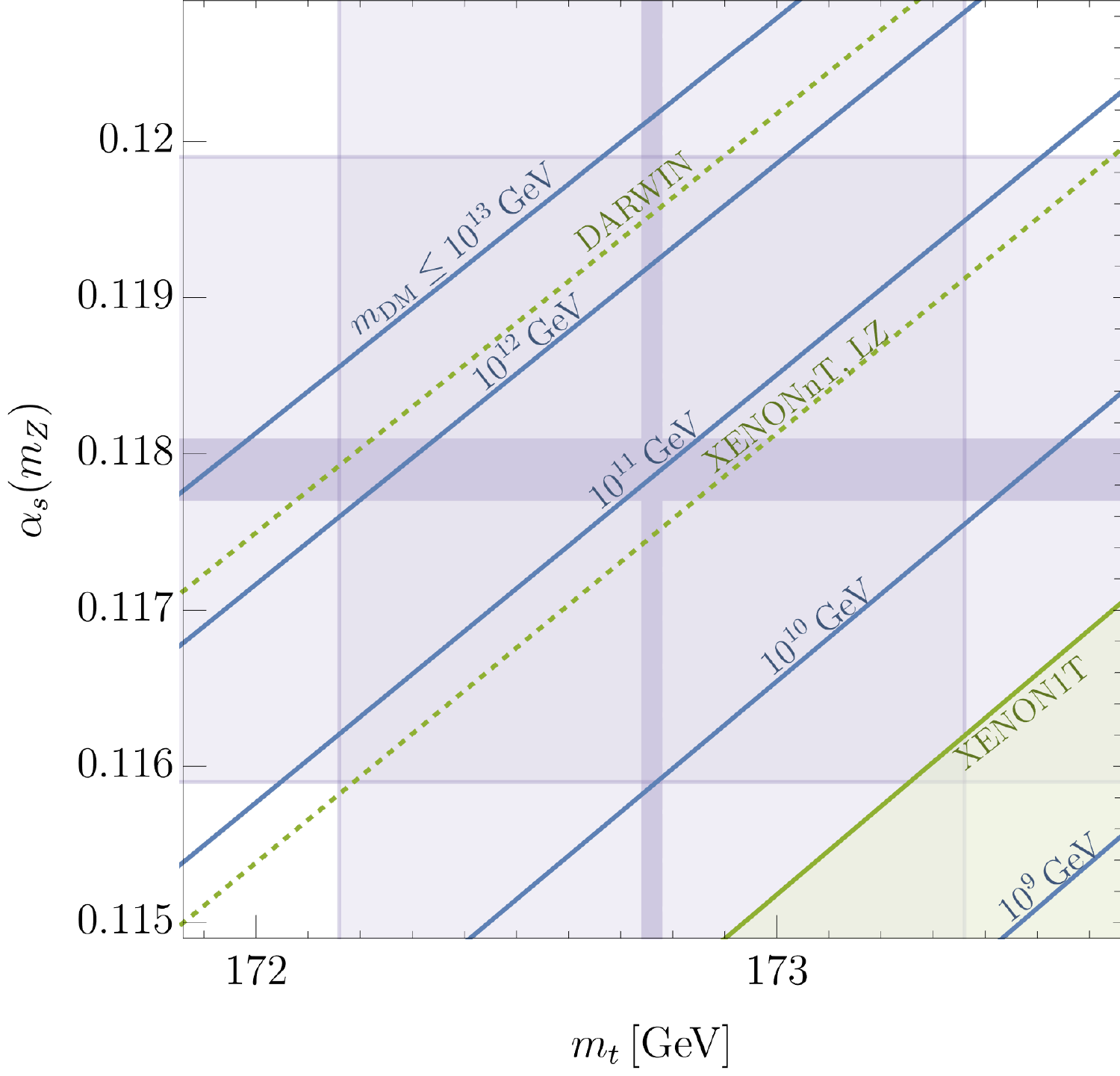} % first figure itself
    %\end{minipage}
    \caption{\small Upper bound on the dark matter mass $m_{\rm DM}$ as a function of the top quark mass $m_{t}$ and the strong coupling constant $\alpha_s(m_Z)$ shown in {\color{c1} \textbf{blue}}. Equivalently, the figure shows an upper bound on $m_t$ as a function of $\alpha_s(m_Z)$ and $m_{\rm DM}$, and a lower bound on $\alpha_s(m_Z)$ as a function of $m_t$ and $m_{\rm DM}$. The wider {\color{gray} \textbf{gray}} bands show the current $2\sigma$ uncertainties of $m_t$ and $\alpha_s(m_Z)$, and the narrower bands show the expected future uncertainties. Dark matter direct detection bounds are shown in {\color{c3} \textbf{green}}.}
    \label{fig:mTopalphasvsmDM}
\end{figure}

In Fig.~\ref{fig:mTopalphasvsmDM}, the upper bound on the dark matter mass or, equivalently, the upper bound on the top quark mass or the lower bound on the strong coupling constant, is shown. Here we impose $\delta\lambda(m_-)>-0.002$. The current $2\sigma$ uncertainty of $m_t$ and $\alpha_s(m_Z)$ are shown by wide bands.
The uncertainty of $\alpha_s(m_Z)$ can be reduced by a factor of $10$ by measurements at future lepton colliders~\cite{Gomez-Ceballos:2013zzn} or improved lattice calculations~\cite{Lepage:2014fla}.
The uncertainty of $m_t$ can be reduced down to few 10 MeV by future lepton colliders~\cite{Seidel:2013sqa,Horiguchi:2013wra,Kiyo:2015ooa,Beneke:2015kwa}. At this stage, the theoretical computation of the running of the Higgs quartic coupling should be improved; the most recent computation~\cite{Buttazzo:2013uya} has a theoretical uncertainty equivalent to the shift of the top quark mass by 100 MeV.

%If the uncertainty is reduced by a factor of $10$ by measurements at future lepton colliders~\cite{Gomez-Ceballos:2013zzn} or improved lattice calculations~\cite{Lepage:2014fla}, the upper bounds become sharper as shown by the narrower thick band.

%The uncertainty of the top quark mass can be reduced down to few 10 MeV by future lepton colliders~\cite{Seidel:2013sqa,Horiguchi:2013wra,Kiyo:2015ooa,Beneke:2015kwa}. At this stage, the theoretical computation of the running of the Higgs quartic coupling should be improved; the up-to-date computation~\cite{Buttazzo:2013uya} has theoretical uncertainty equivalent to the shift of the top quark mass by 100 MeV.

%%%%%%
%%%%%%
%%%%%%
%%%%%%

%%%%%%
%%%%%%
%%%%%%
%%%%%%

\section{Supersymmetry Breaking Constrained by Unification }
\label{sec:unifiedsusyBC}
In this section, we discuss the quartic coupling at the supersymmetry breaking scale, $\tilde{m}$, starting from boundary conditions at the unification scale $\sim 10^{16}$ GeV. We show that the tree-level quartic coupling is typically $0.001-0.01$.

\begin{figure}[tb]
    \begin{center}
           \includegraphics[width=0.6\textwidth]{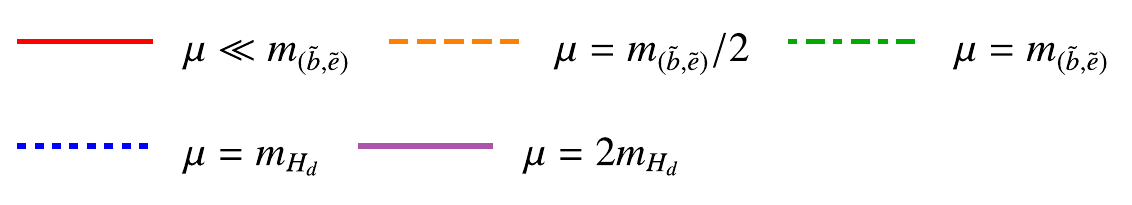}

        \includegraphics[width=0.49\textwidth]{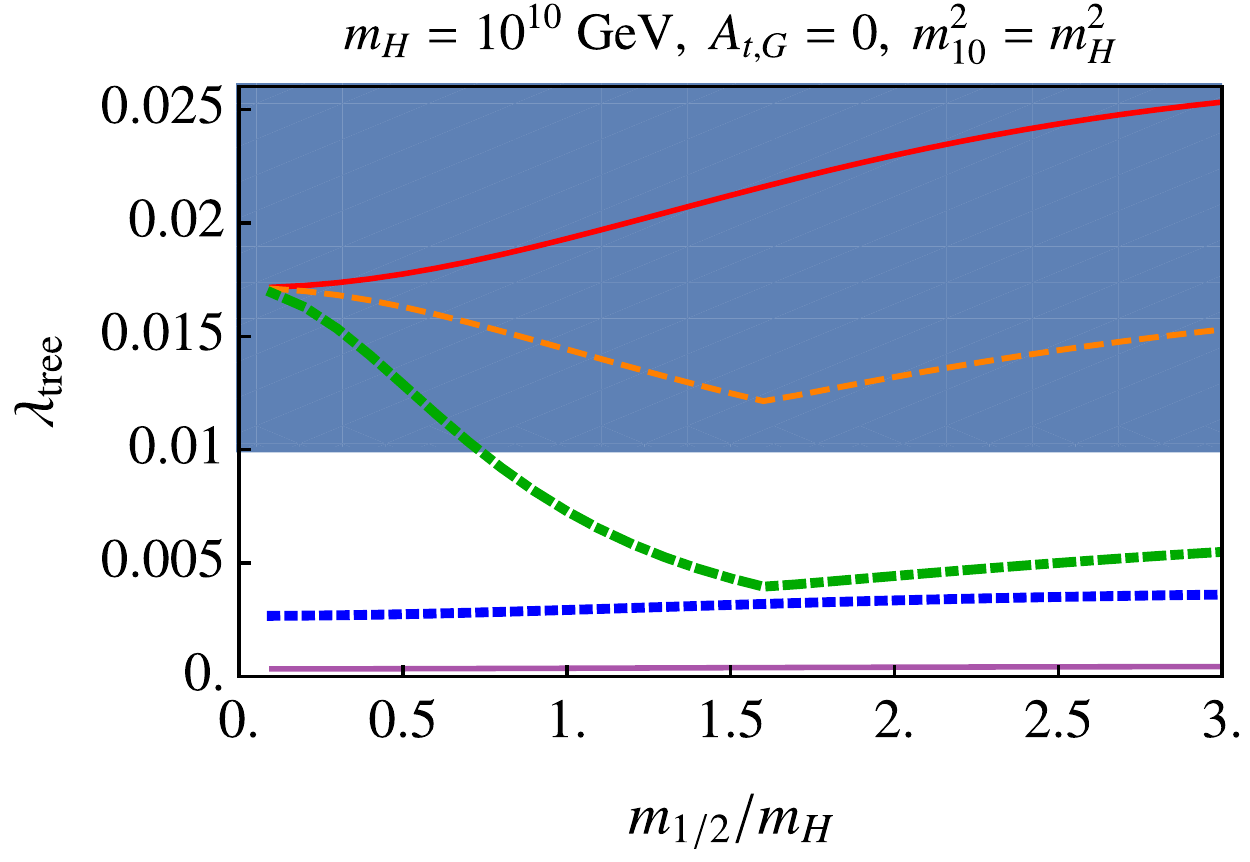}
         \includegraphics[width=0.49\textwidth]{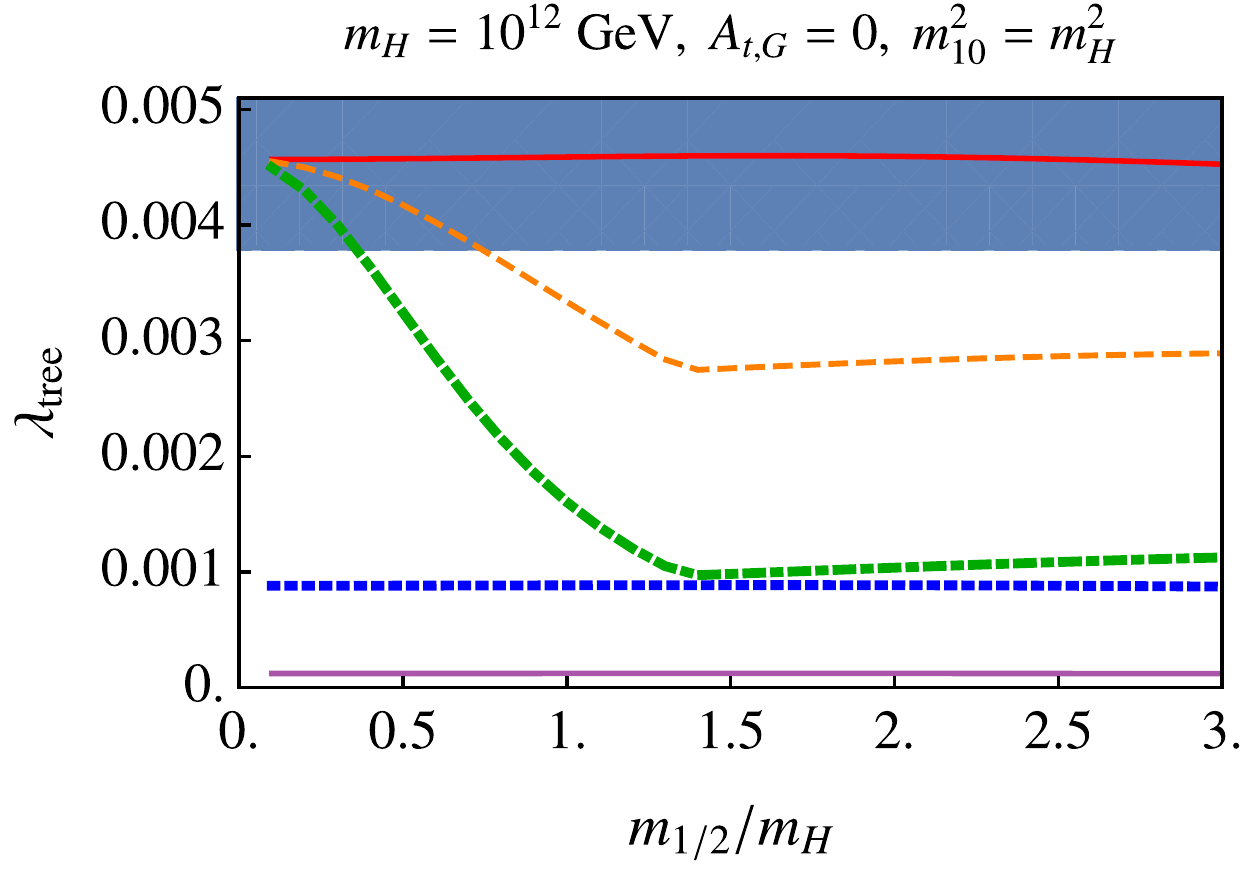}
         \includegraphics[width=0.49\textwidth]{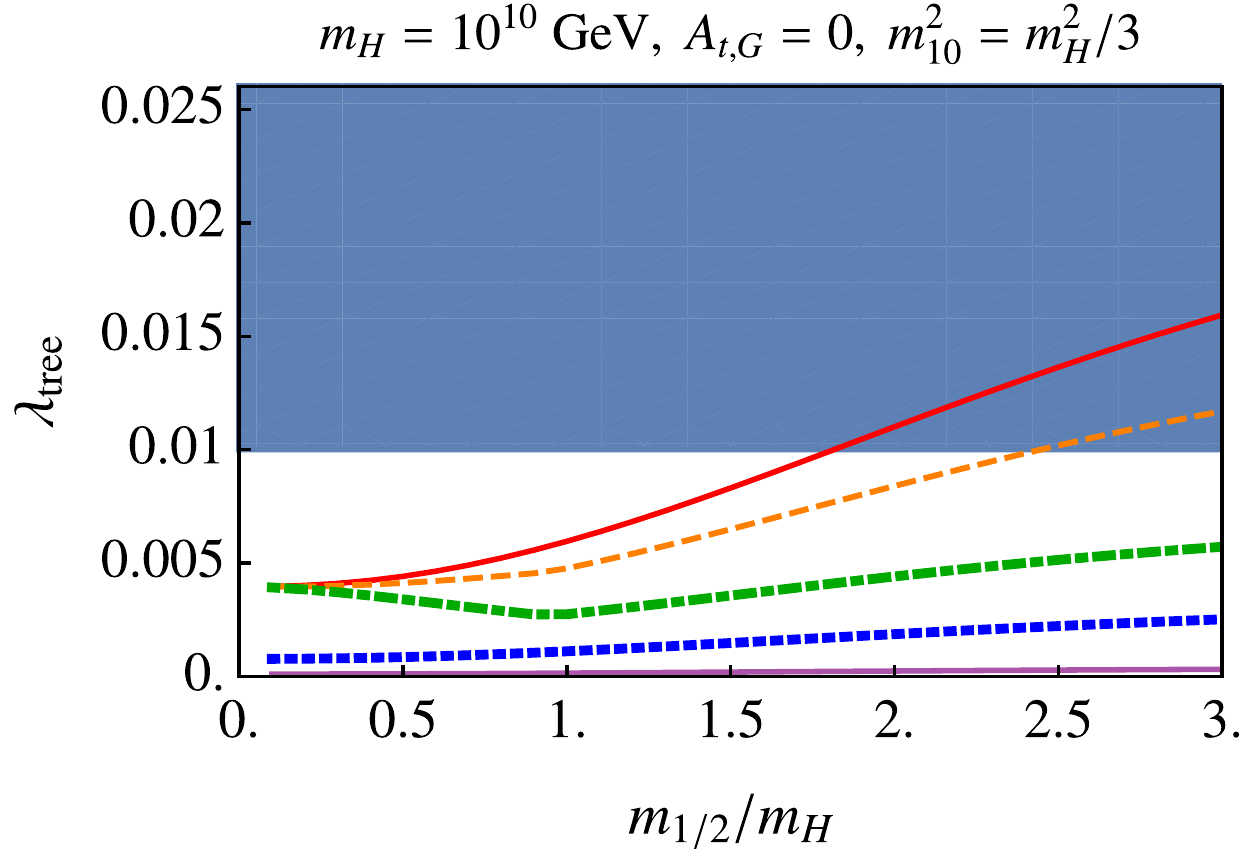}
         \includegraphics[width=0.49\textwidth]{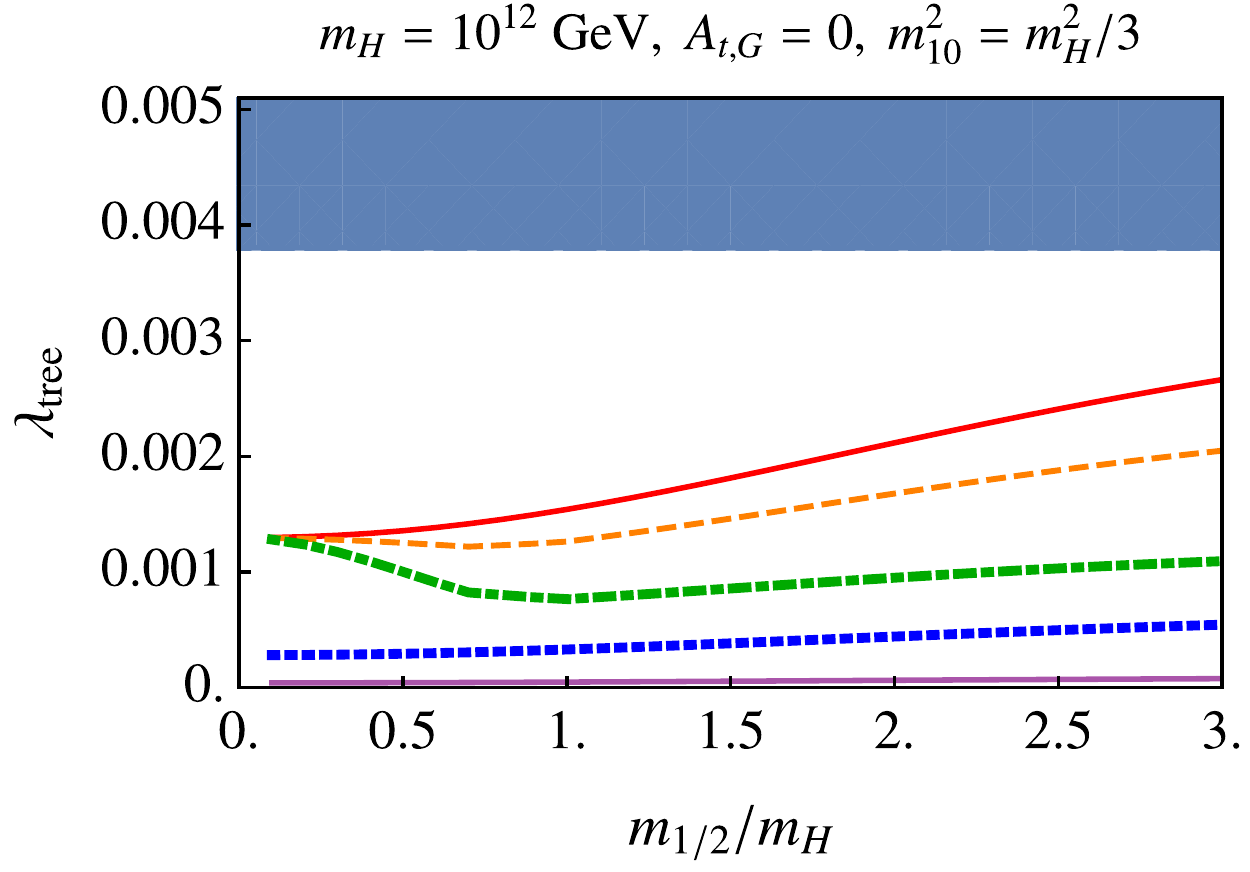}
          \includegraphics[width=0.49\textwidth]{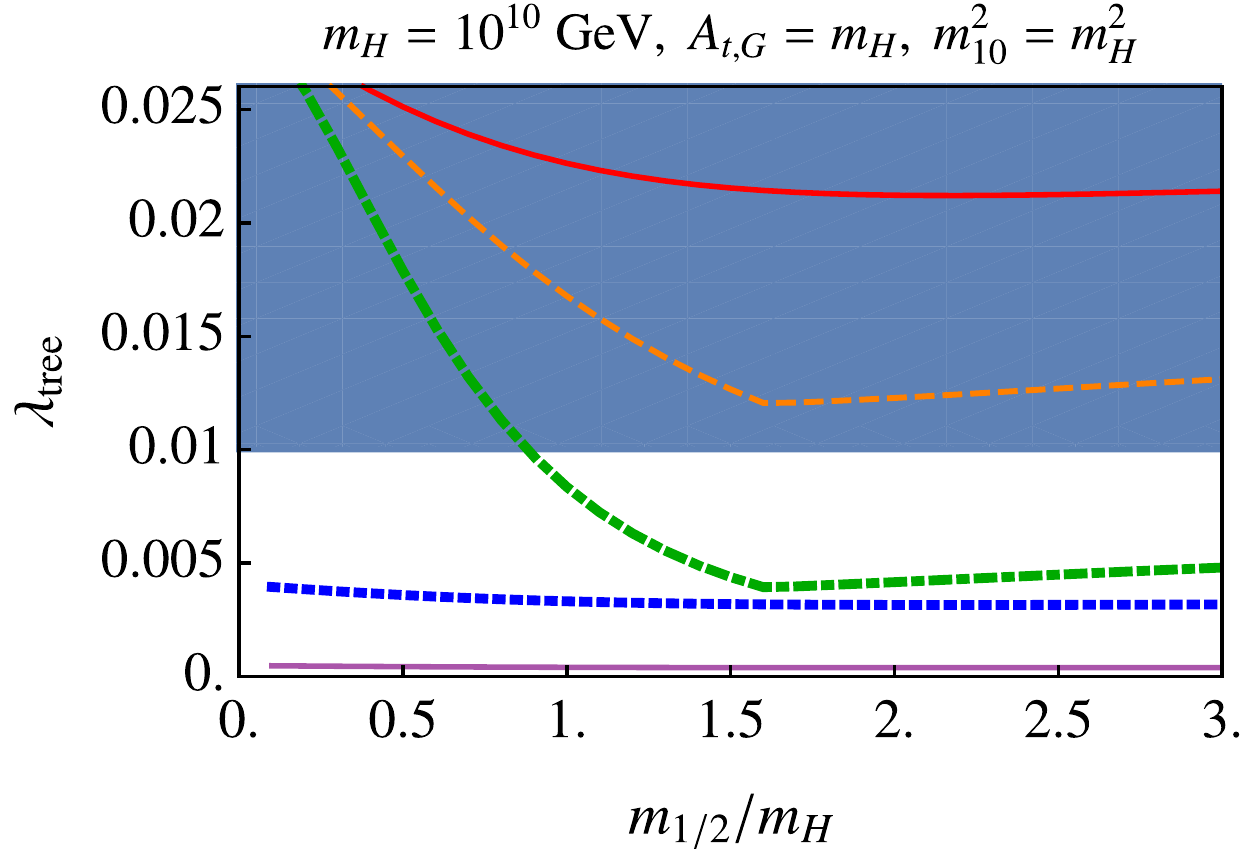}
          \includegraphics[width=0.49\textwidth]{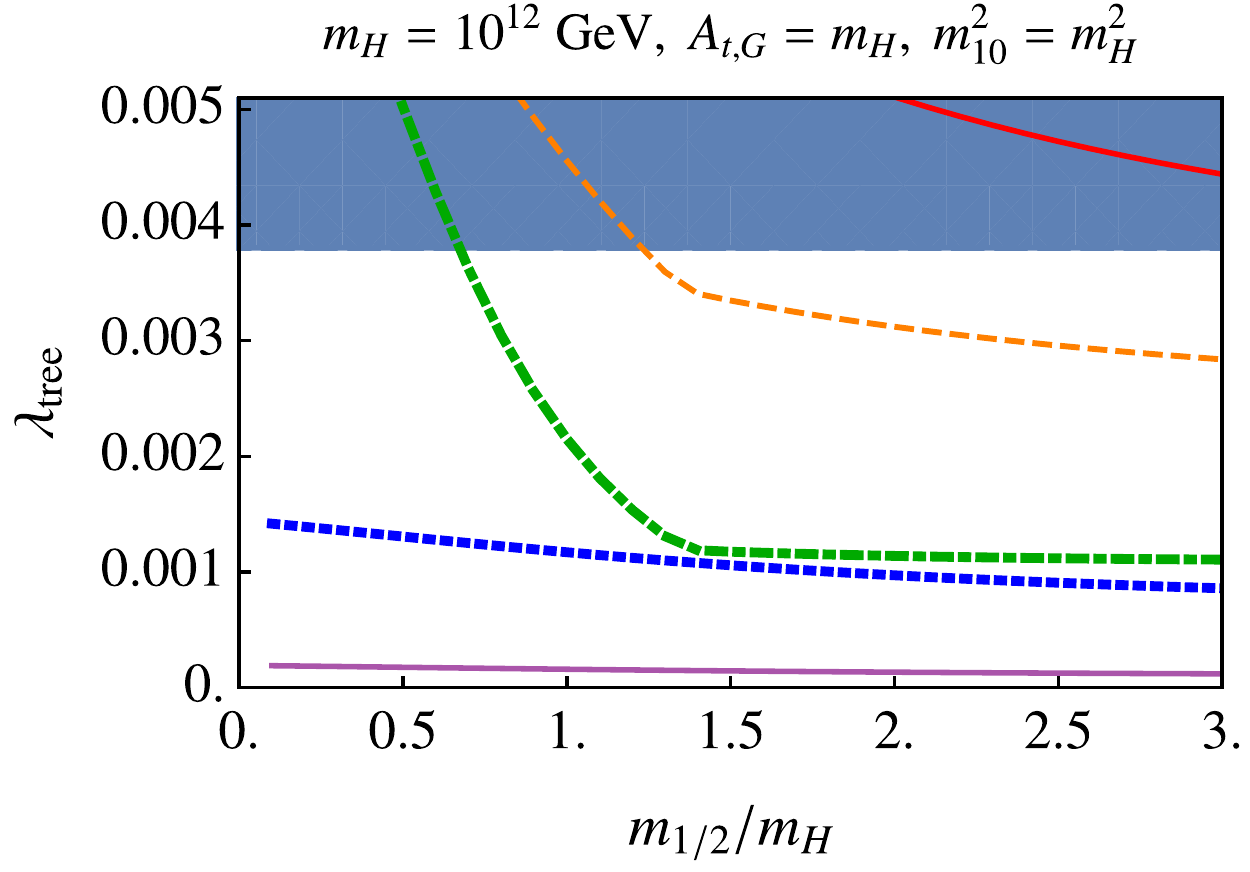}
    \end{center}
     \caption{Prediction for the tree-level quartic coupling with a UV boundary condition $m_{H_u}=m_{H_d}$. In the {\color{c1} \textbf{blue}} shaded region, reproducing $\lambda(m_-)$ requires $m_t < 171.86$ GeV, $3\sigma$ away from the central value. Here we impose $\alpha_s(m_Z)<0.1189$ and $\delta\lambda(m_-)> - 0.002$.}
    \label{fig:quartic_UV}    
\end{figure}

As we have seen, the quartic coupling at $\tilde{m}$ is small when $m_{H_u}^2 \sim m_{H_d}^2$. A relation $m_{H_u}^2 = m_{H_d}^2$ can be naturally realized at a high energy scale by a symmetry relating $H_u$ with $H_d$, such as a discrete symmetry or $SO(10)$ gauge symmetry, or a universality of scalar masses. The relation is necessarily destabilized by quantum correction from the top quark Yukawa coupling,
\begin{align}
    \frac{\rm d}{{\rm dln}\mu} m_{H_u}^2 = \frac{3y_t^2}{8\pi^2} \left( m_{H_u}^2 + m_{\tilde{q}}^2 + m_{\tilde{\bar{u}}}^2  + A_t^2 \right) + \cdots,
\end{align}
where the ellipsis denotes terms independent of the top Yukawa.
We compute the renormalization group running of the MSSM from a scale $10^{16}$ GeV down to $\tilde{m}$ with a UV boundary condition motivated from $SU(5)$ unification,
\begin{align}
    m_{H_u}^2 = m_{H_d}^2 = m_{H}^2,~~m_{\tilde{q}}^2 = m_{\tilde{\bar{u}}}^2 = m_{\tilde{\bar{e}}}^2 = m_{10}^2,~~m_{\tilde{\bar{d}}}^2 = m_{\tilde{\ell}}^2 = m_{5}^2,\nonumber \\
    M_{1}= M_2= M_3 = m_{1/2},~~A_t = A_{t,G}.
\end{align}
The SM top yukawa coupling is matched to the MSSM top yukawa coupling at $\tilde{m}$ assuming ${\rm tan}\beta \simeq 1$, $y_{t,{\rm MSSM}} = \sqrt{2}y_{t,{\rm SM}}$.
The soft masses $m_{H_u}^2$ and $m_{H_d}^2$ at the renormalization scale $(10^{10}, 10^{12})$ GeV are given by the  analytic results
\begin{align}
    m_{H_u}^2(10^{10}~{\rm GeV}) =& 0.77 m_H^2 &-0.46 m_{10}^2& - 0.12 A_{t,G}^2& + 0.02 m_{1/2}^2& + 0.08 m_{1/2} A_{t,G},\nonumber \\
    m_{H_d}^2(10^{10}~{\rm GeV}) =& 1.0 m_H^2 &&& + 0.19 m_{1/2}^2,
\end{align}
\begin{align}
    m_{H_u}^2(10^{12}~{\rm GeV}) =& 0.86 m_H^2 &-0.28 m_{10}^2& - 0.10 A_{t,G}^2& + 0.06 m_{1/2}^2& + 0.04 m_{1/2} A_{t,G}, \nonumber\\
    m_{H_d}^2(10^{12}~{\rm GeV}) =& 1.0 m_H^2 &&& + 0.12 m_{1/2}^2.
\end{align}

In Fig.~\ref{fig:quartic_UV}, we show the tree-level quartic coupling as a function of $m_{1/2}/m_H$ for several representative boundary conditions; the left (right) panels have $m_{H}=10^{10}$ GeV ($10^{12}$ GeV).
We fix the renormalization scale to be the matching scale used in the previous section, $m_-$, the lighter of $m_{\tilde{q}}$ and $m_{\tilde{\bar{u}}}$. The boundary condition for $m_5^2$ is not specified as it does not affect the renormalization group running of $m_{H_u}$.
Note that the bino, $\tilde{b}$, is the lightest gaugino and the right-handed slepton, $\tilde{e}$, is the lightest scalar in the matter ten-plet.
We define $m_{(\tilde{b}, \tilde{e})}$ to be the smaller of $m_{\tilde{b}}$ and $m_{\tilde{e}}$.  On the five lines, $\mu$ is fixed to be $(\ll m_{(\tilde{b}, \tilde{e})},\; m_{(\tilde{b}, \tilde{e})}/2,\; m_{(\tilde{b}, \tilde{e})},\;  m_{H_d},\; 2m_{H_d})$. As $\mu$ is increased, the tree-level quartic coupling decreases rapidly, as expected from (\ref{eq:treeQuartic}), (\ref{eq:cossq2beta}) and Fig.~\ref{fig:λISSPlot}.
For large values of $m_5^2$ the Higgsino is the LSP above the green dot-dashed line, and the region below the line is excluded because at low (high) $m_{1/2}$ the LSP is the bino (a charged right-handed slepton). For small values of $m_5^2$ the tau sneutrino can be the LSP throughout the plane, although at low $\mu$ the Higgsino LSP is also possible. 
In the blue shaded region, the top quark mass must be below $171.86$ GeV, more than $3\sigma$ away from the central value, in order for $\lambda(m_-)$ to be consistent with the running of the Higgs quartic coupling. To derive a conservative bound, we take $\alpha_s(m_Z)= 0.1189$, $1\sigma$ above the central value, and $\delta \lambda = - 0.002$, the smallest realistic threshold correction.

We see that smaller values of $\lambda_{{\rm tree}}$ result for larger $m_H$, which gives less running, larger values of $\mu/m_H$ and smaller values of $m_{10}/m_H$ and $A_{t,G}/m_H$. For $m_H = 10^{12}$ GeV, $\lambda_{{\rm tree}} < 0.003$ over much of the parameter space.  Including threshold corrections, Fig.~\ref{fig:λvsmtildeLoopPlot} shows that this is ideal for consistency with the observed Higgs mass, and requires a low value of the top quark mass.  For $m_H = 10^{10}$ GeV, $\lambda_{{\rm tree}} < 0.01$ over much of the parameter space, except at low values of $\mu$, which from Fig.~\ref{fig:λvsmtildeLoopPlot} again shows excellent consistency with the observed Higgs mass, and leads to the expectation that Higgsino/sneutrino dark matter will be discovered at planned experiments.

%For small/large $m_{1/2}$, the NLSP is the bino/right-handed stau.  but we specify its value since the lightest sfermion becomes the sneutrino if $m_5 < m_{10}$ and affects the predictions of the tree-level quartic coupling for $\mu = m_{\rm NLSP}$ and $m_{\rm NLSP}/2$.  For $\mu = (2)m_{H_d}$, the Higgsino cannot be the LSP and hence the sneutrino should be the LSP for observable direct detection signals in near future. This can be achieved by taking $m_5$ smaller than $m_{10}$.

 %Above the blue line with $\mu = m_{NLSP}$ the Higgsino is the LSP. Below this line we require the tau sneutrino to be the LSP, which occurs when $m_{1/2}$ is sufficiently large to avoid bino LSP. Here, $m_5$ does not affect the renormalization group running of $m_{H_u}$ but we specify its value since the lightest sfermion becomes the sneutrino if $m_5 < m_{10}$ and affects the predictions of the tree-level quartic coupling for $\mu = m_{\rm NLSP}$ and $m_{\rm NLSP}/2$.  For $\mu = (2)m_{H_d}$, the Higgsino cannot be the LSP and hence the sneutrino should be the LSP for observable direct detection signals in near future. This can be achieved by taking $m_5$ smaller than $m_{10}$.

\section{Cosmological Abundance of LSP with Intermediate Scale Mass}
\label{sec:cosabund}

In this section, we discuss how the heavy LSP dark matter can be populated in the early universe. Most of the discussion in this section is applicable to generic heavy dark matter with electroweak interactions.
Standard freeze-out during the radiation dominated era overproduces the LSP because of its large mass. To avoid this, the reheating temperature of the universe must be smaller than the LSP mass, and the LSP must be produced during the reheating process. We discuss reheating by the inflaton $\phi$, but, if the LSPs produced during inflaton reheating are subdominant, the following discussion also applies to the case where some other particle or condensate dominates the energy density of the universe. 

\subsection{Direct decay of the inflaton}
The inflaton can directly decay into sparticles if its  mass is more than double the LSP mass. The energy density of the LSP normalized by the entropy density is
\begin{align}
    \frac{\rho_{\rm LSP}}{s} \simeq N_{\rm LSP}\frac{ m_{\rm DM}T_{\rm RH}}{m_\phi} = 10^3~{\rm eV} \frac{m_{\rm DM}}{10^{10}~{\rm GeV}} \frac{10^{13}~{\rm GeV}}{m_{\phi}} \frac{T_{\rm RH}}{\rm MeV} N_{\rm LSP},
\end{align}
where $N_{\rm LSP}$ is the number of LSPs produced per inflaton decay. Because of supersymmetry, $N_{\rm LSP}$ is at the smallest $O(1)$. When $m_{\phi} \gg m_{\rm DM}$ and the inflaton dominantly decays into SM charged particles, showering leads to $N_{\rm LSP} \gg 1$~\cite{Kurata:2012nf,Harigaya:2016vda}. Giving the lower bound $T_{\rm RH}>4$ MeV~\cite{Kawasaki:1999na,Kawasaki:2000en,Hasegawa:2019jsa}, it is difficult to produce the correct LSP abundance in this way. Hence, the inflaton mass must be below the sparticle mass scale.
(In this case, production of the LSP via scattering among the inflaton decay products and the thermal bath~\cite{Harigaya:2014waa,Garcia:2018wtq,Harigaya:2019tzu} is absent.)

\subsection{Production during the inflaton dominated era}
We first derive the evolution of the temperature of the universe. We consider the case where the dissipation of the inflaton occurs by perturbative processes, with dissipation rates given by
\begin{align}
    \Gamma =
    \begin{cases}
    \Gamma_0 & : T<m_\phi \\
    \Gamma_0 \left(\frac{T}{m_\phi}\right)^n & : m_\phi <T
    \end{cases}.
\end{align}
For $T<m_\phi$, dissipation is governed by the zero-temperature decay rate $\Gamma_0$, while for $m_\phi<T$, thermal effects should be taken into account. $n=1$ arises when dissipation is caused by a dimensionless coupling, while $n=-1$ arises when dissipation is caused by a dimension-3 coupling, such as $\phi h h^\dag$.

The dependence of the temperature on the Hubble scale is given by
\begin{align}
\label{eq:T1}
 T_{\rm RH}< m_\phi~~ &:~~   T =
    \begin{cases}
    T_{\rm RH} \left(\frac{H}{H_{\rm RH}}\right)^{1/4} & : T_{\rm RH} < T < m_\phi \\
    m_\phi \left( \frac{H T_{\rm RH}^4}{H_{\rm RH} m_\phi^4}\right)^{1/(4-n)} & : m_\phi <T,
    \end{cases} \\
\label{eq:T2}
     m_\phi < T_{\rm RH} ~~ &: ~~ T = T_{\rm RH} \left( \frac{H }{H_{\rm RH}}\right)^{1/(4-n)},
\end{align}
where $H_{\rm RH} = \sqrt{\pi^2 g_*/90} \; T_{\rm RH}^2/\Mpl$ is the Hubble scale at the completion of reheating.
We implicitly assumed that the radiation is thermalized, which is not satisfied for small $T_{\rm RH}$ and/or large $T$. We discuss thermalization while discussing the production of the LSP below.

\subsubsection*{Case 1: $T_{\rm FO}< m_\phi <2 m_{\rm DM}$}
During freeze-out, when $T_{\rm FO} = m_\phi/x_{\rm FO} < m_\phi$,  radiation is created from the zero-temperature decay of the inflaton and the temperature of the universe is given by the first line of Eq.~(\ref{eq:T1}). Such a case is studied in the literature assuming efficient thermalization~\cite{Chung:1998rq,Giudice:2000ex}.

After freeze-out, the LSP number density, normalized by the inflaton energy density,  is
\begin{align}
    \frac{n_{\rm LSP}}{\rho_\phi} \simeq \frac{H_{\rm FO}}{\vev{\sigma v} \rho_\phi} = \frac{1}{3 \vev{\sigma v}H_{\rm FO}\Mpl^2}.
\end{align}
Using $\rho_\phi/s \simeq 3T_{\rm RH}/4$ at the completion of reheating, we obtain
\begin{align}
\label{eq:DM1}
    \frac{\rho_{\rm LSP}}{s} \simeq \frac{x_{\rm FO}^{4}}{4} \sqrt{\frac{90}{\pi^2 g_*}} \frac{1}{4\pi \alpha_2^2} \frac{T_{\rm RH}^3}{\Mpl m_{\rm DM}} \frac{4\pi \alpha_2^2 / m_{\rm DM}^2}{\vev{\sigma v}}.
\end{align}

Here, we assume that radiation thermalizes around the freeze-out temperature. This assumption is valid if $4\pi \alpha^2 T_{\rm FO} > H_{\rm FO}$, requiring
\begin{align}
    T_{\rm RH} > \left[ \frac{1}{4\alpha^2 }\sqrt{\frac{g_*}{90}} \frac{(m_{\rm DM}/x_{\rm FO})^3}{ \Mpl} \right]^{1/2} \equiv T_{\rm RH,th}.
\end{align}
If this condition is violated, the radiation produced from the inflaton does not reach thermal equilibrium by the would-be freeze-out.
We expect that the distribution of radiation in this case is close to that after preheating~\cite{Micha:2002ey,Micha:2004bv}.
Since scattering is efficient at lower energies, the lower energy modes are populated. The typical energy of the radiation is below the would-be temperature and the radiation is in an over-occupied state. The energy distribution has a cutoff, above which the scattering is inefficient and the distribution is exponentially suppressed.

For large $m_{\rm DM}$, the reheating temperature to reproduce the observed abundance from Eq.~(\ref{eq:DM1}) is in fact smaller than $T_{\rm RH,th}$. Then the LSP abundance is exponentially suppressed and LSPs are under-produced. As $T_{\rm RH}$ approaches $T_{\rm RH,th}$, the LSP production is not suppressed, and the freeze-out picture is applicable. Since $T_{\rm RH}\sim T_{\rm RH,th}$ is larger than that to produce an appropriate amount of LSPs  according to Eq.~(\ref{eq:DM1}), LSPs are over-produced. Thus, the observed dark matter abundance can be reproduced for $T_{\rm RH}$ slightly below $T_{\rm RH,th}$. We call this scenario {\it non-thermal freeze-in}.

\begin{figure}[tb]
    \begin{center}
     \includegraphics[width=0.49\textwidth]{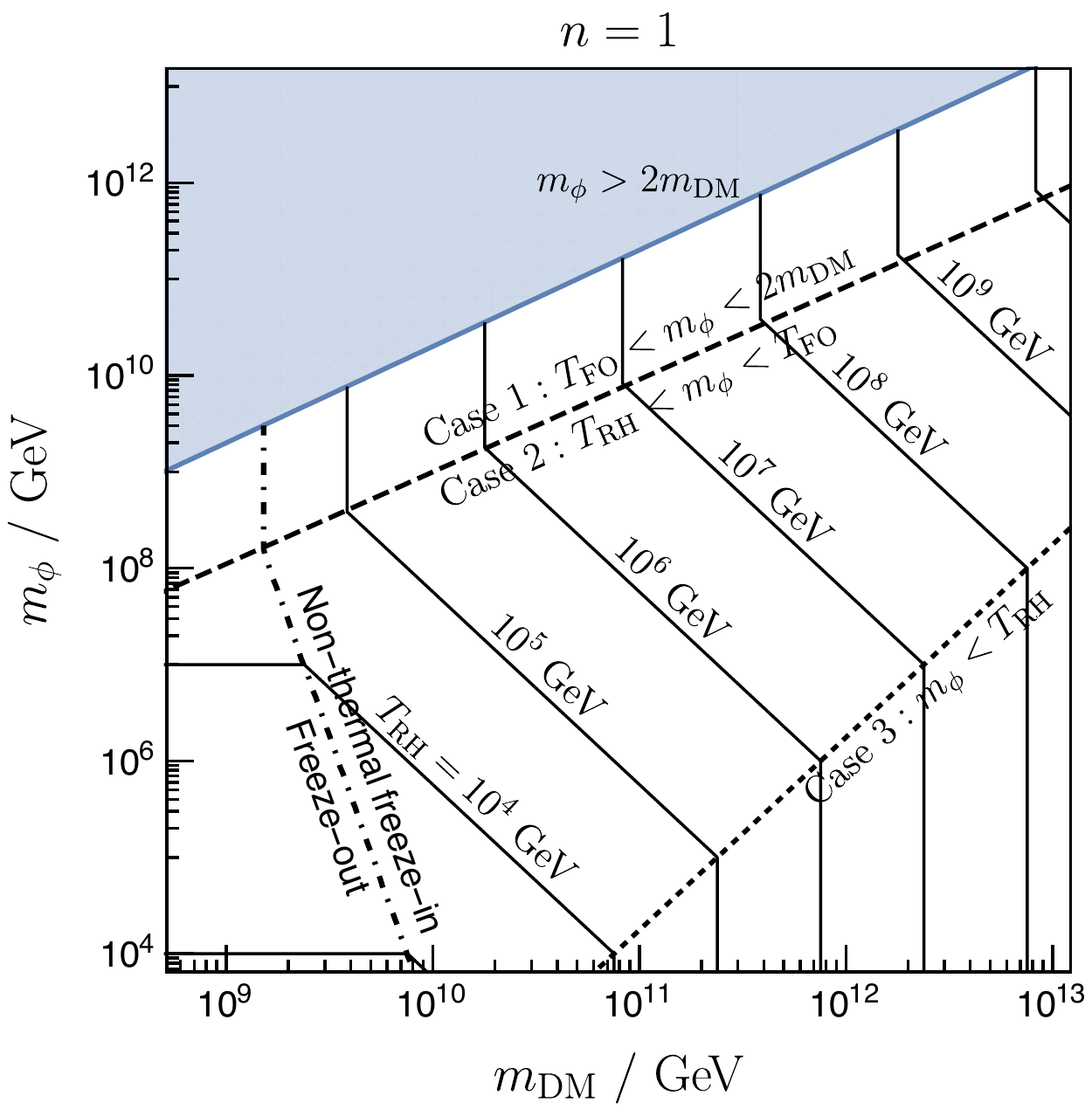}
        \includegraphics[width=0.49\textwidth]{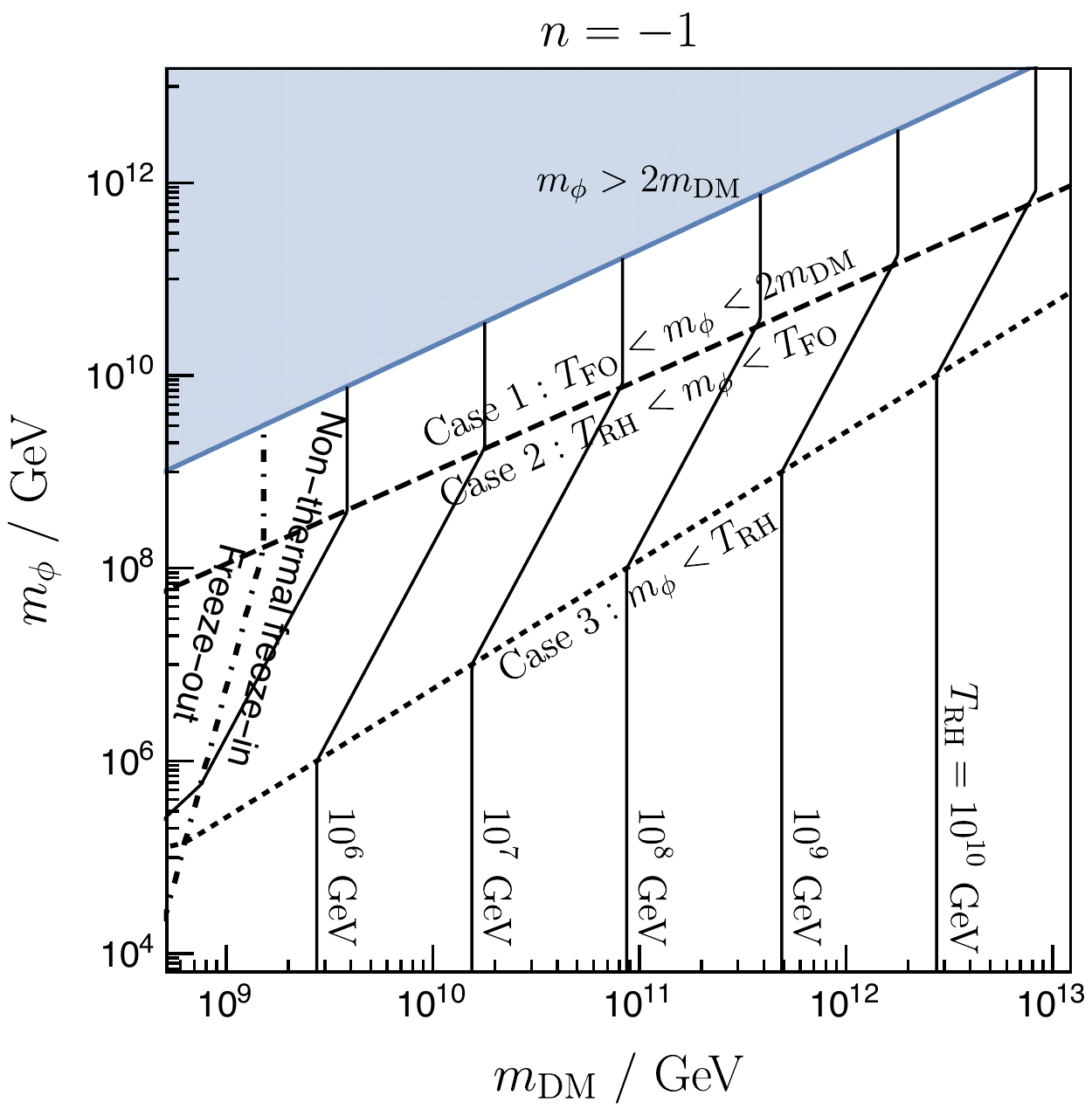}
    \end{center}
    \caption{Contours of the reheating temperature $T_{\rm RH}$ required to produce the observed dark matter abundance by LSP production during reheating. In the {\color{c1} \textbf{blue}} region, direct decay of the inflaton into sparticles overproduces the LSP. To the right of the dot-dashed line, radiation is not thermalized by the would-be freeze-out, and the LSP production occurs just before the completion of thermalization.}
    \label{fig:TRH}    
\end{figure}

The required reheating temperature to produce the observed dark matter abundance by LSP production during reheating is shown in Fig.~\ref{fig:TRH}. Above the black dashed line, $T_{\rm FO} < m_\phi < 2m_{\rm DM}$ and the analysis shown above is applicable. To the left of the black dot-dashed line, the LSP abundance is determined by freeze-out, while to the right, the abundance is determined by the exponentially suppressed production just before thermalization.

\subsubsection*{Case 2: $T_{\rm RH}<m_\phi<T_{\rm FO}$}
For the inflaton mass between $T_{\rm RH}$ and $T_{\rm FO}$, the temperature of the universe during freeze-out is given by the second line of Eq.~(\ref{eq:T1}). By a computation similar to that which leads to Eq.~(\ref{eq:DM1}), we obtain 
\begin{align}
\label{eq:DM2}
    \frac{\rho_{\rm LSP}}{s} \simeq \frac{x_{\rm FO}^{4-n}}{4} \sqrt{\frac{90}{\pi^2 g_*}} \frac{1}{4\pi \alpha_2^2} \frac{T_{\rm RH}^3}{\Mpl m_\phi^n m_{\rm DM}^{1-n}} \frac{4\pi \alpha_2^2 / m_{\rm DM}^2}{\vev{\sigma v}}.
\end{align}
Radiation thermalizes before freeze-out if
\begin{align}
     T_{\rm RH} > \left[ \frac{1}{4\alpha^2 }\sqrt{\frac{g_*}{90}} \frac{(m_{\rm DM}/x_{\rm FO})^{3-n} m_\phi^{n} }{  \Mpl} \right]^{1/2}\equiv T_{\rm RH,th}.
\end{align}
The reheating temperature required to produce the observed dark matter abundance is shown in Fig.~\ref{fig:TRH}. The above analysis is applicable between the dashed and dotted lines.

\subsubsection*{Case 3: $m_\phi<T_{\rm RH}$}
For the inflaton mass below the reheating temperature, the temperature during freeze-out is given by Eq.~(\ref{eq:T2}).
The LSP density is given by
\begin{align}
    \frac{\rho_{\rm LSP}}{s} \simeq \frac{x_{\rm FO}^{4-n}}{4} \sqrt{\frac{90}{\pi^2 g_*}} \frac{1}{4\pi \alpha_2^2} \frac{T_{\rm RH}^{3-n}}{\Mpl m_{\rm DM}^{1-n}} \frac{4\pi \alpha_2^2 / m_{\rm DM}^2}{\vev{\sigma v}}.
\end{align}
Radiation thermalizes before freeze-out if
\begin{align}
    T_{\rm RH} > \left[ \frac{1}{4 \alpha^2} \sqrt{\frac{g_*}{90}} \frac{(m_{\rm DM}/x_{\rm FO})^{3-n}}{\Mpl} \right]^{\frac{1}{2-n}} \equiv T_{\rm RH,th}.
\end{align}
The reheating temperature required to produce the observed dark matter abundance is shown in Fig.~\ref{fig:TRH}. This analysis is applicable below the dotted line.

\subsection{Other possibilities}
Is is possible that the maximal temperature of the universe is the reheating temperature. This occurs when reheating is instantaneous, the dissipation rate of the inflaton increases towards the end of inflation~\cite{Co:2020xaf}, or a kinematically available decay channel opens suddenly~\cite{Fujita:2016vfj}.
In this case, the correct LSP abundance is obtained if the reheating temperature is about $m_{\rm DM}/10$, so that the LSP production is exponentially suppressed.

The evolution of the early universe may include an era of domination by primordial black holes (PBHs). If the initial Hawking temperature of the PBHs is below $m_{\rm DM}$, the PBHs emit LSPs only after they lose most of their mass by Hawking radiation into light particles, and the LSP abundance is suppressed. As a result the correct LSP abundance can be obtained for sufficiently large initial PBH masses~\cite{Green:1999yh,Fujita:2014hha}.

\section{Conclusions}
\label{sec:conclusions}

In recent decades, the theoretical and experimental investigations of supersymmetry were focused on weak scale supersymmetry. The discovery of the Higgs with a mass of $125$ GeV has revealed a new scale of the SM, the Higgs quartic scale $\mu_\lambda=10^{9-12}$ GeV, at which the SM Higgs quartic coupling vanishes.
In this paper, we focused on Intermediate Scale Supersymmetry where supersymmetry is broken near the Higgs quartic scale. In this framework, including threshold corrections we found a small SM Higgs quartic coupling for a wide range of supersymmetry breaking parameters.
The LSP is a dark matter candidate, and we studied the cases of Higgsino and sneutrino LSP, which scatter with nuclei via tree-level $Z$ boson exchange. Direct detection experiments have already excluded the LSP mass below $3 \times 10^9$ GeV, and will probe it up to $10^{12}$ GeV.

The Higgs quartic scale is sensitive to SM parameters. Currently, the uncertainty of the scale is dominated by the top quark mass and the strong coupling constant. We derived an upper bound on the LSP mass as a function of the top quark mass and the strong coupling constant shown in Fig.~\ref{fig:mTopalphasvsmDM}. Around the central value of SM parameters, dark matter signals should be discovered by near future experiments.
Conversely, the figure shows an upper bound on the top quark mass and a lower bound on the strong coupling constant as a function of the LSP mass.

We also discussed how this LSP dark matter may be populated in the early universe. Because of the large LSP mass, the standard freeze-out mechanism overproduces the LSP. We avoid this by taking the reheating temperature after inflation below the LSP mass.  We find that the observed dark matter abundance can be obtained during the reheating era, and in most of the parameter space, the inflaton condensate is dissipated by thermal effects during LSP production. We determined the required reheating temperature as a function of the inflaton mass and the LSP mass. Once the LSP mass is fixed by the signal rate at direct detection experiments, the reheating temperature is predicted from the inflaton mass.

\section*{Acknowledgement}
This work was supported in part by the Director, Office of Science, Office of High Energy and Nuclear Physics, of the US Department of Energy under Contracts DE-AC02-05CH11231 (LJH), by the National Science Foundation under grant PHY-1915314 (LJH), as well as by Friends of the Institute for Advanced Study (KH).

\appendix

\section{Stability bound on a trilinear coupling}
\label{app:instability}
In this appendix, we derive an upper bound on the trilinear coupling between between the Higgs and stops from the stability of the electroweak vacuum. We consider the case of ${\rm tan}\beta \simeq 1$ and field directions parameterized by
\begin{align}
H_u \rightarrow \frac{1}{2}
\begin{pmatrix}
h - H \\
0
\end{pmatrix},~
H_d \rightarrow \frac{1}{2}
\begin{pmatrix}
 0\\
h + H
\end{pmatrix},~~
q \rightarrow \frac{1}{\sqrt{2}} 
\begin{pmatrix}
u & 0 \\
0 & 0 \\
0 & 0
\end{pmatrix},~~
\bar{u} \rightarrow \frac{1}{\sqrt{2}} 
\begin{pmatrix}
\bar{u}  \\
0  \\
0 
\end{pmatrix},
\end{align}
where $h$, $H$, $u$, and $\bar{u}$ are real fields with potential
\begin{align}
    V(h,H,q,\bar{u}) = & \frac{1}{2}m_A^2H^2 + \frac{1}{2}m_{\tilde{q}}^2 q^2 + \frac{1}{2} m_{\tilde{u}}^2 u^2  - \frac{1}{\sqrt{2}}y_t h(A-\mu) u \bar{u}  -  \frac{1}{\sqrt{2}}y_t H(A+\mu)u\bar{u} \nonumber \\
    & + y_t^2 \left( \frac{1}{2} u^2 \bar{u}^2 + \frac{1}{4} \left( h + H\right)^2 \left(u^2 + \bar{u}^2\right) \right) + \frac{g^{'2}}{2} \left( \frac{1}{2} h H + \frac{1}{12} u^2 - \frac{1}{3} \bar{u}^2\right)^2 \nonumber \\
   & + \frac{g^2}{2} \left( \frac{1}{2} h H - \frac{1}{4} u^2 \right)^2
    + \frac{g_3^2}{24} \left(u^2 - \bar{u}^2\right)^2.
\end{align}
The renormalization scale of the coupling constants is taken to be around the sparticle mass scale.

The tunneling rate per volume is given by~\cite{Coleman:1977py}
\begin{align}
 \frac{\Gamma}{V} = M^4 {\rm exp}^{-S_B},
\end{align}
where $S_B$ is a bounce action and $M$ is a pre-factor as large as the typical energy scale associated with the tunneling, which we take to be the sparticle mass scale. To avoid tunneling into another vacuum, we require that $\Gamma/V \times H_0^4<1$. For sparticle masses around $10^{10}$ GeV, this corresponds to
\begin{align}
    S_B < 480.
\end{align}

We computed the bounce action using SimpleBounce~\cite{Sato:2019wpo}.
For $m_{\tilde{q}}= m_{\tilde{u}} = m_A \equiv \tilde{m} $ and $A+\mu = 0$, we obtain
\begin{align}
    |A-\mu| < (3.2-3.3) \tilde{m}
\end{align}
for $\tilde{m} = (10^{10}-10^{12})$ GeV.  
The upper bound excludes large values of $A-\mu$ that would give a negative threshold correction to $\lambda$. For $A+\mu \neq 0$, the bound becomes stronger. Larger $m_A$ slightly relaxes the bound, but not enough to enable a negative threshold correction to $\lambda$ from the trilinear coupling.

\bibliographystyle{JHEP}
\bibliography{DMDDISS}

\providecommand{\href}[2]{#2}\begingroup\raggedright\begin{thebibliography}{10}

\bibitem{Goodman:1984dc}
M.~W. Goodman and E.~Witten, \emph{{Detectability of Certain Dark Matter
  Candidates}}, \href{https://doi.org/10.1103/PhysRevD.31.3059}{\emph{Phys.
  Rev. D} {\bfseries 31} (1985) 3059}.

\bibitem{Aprile:2018dbl}
{\scshape XENON} collaboration, \emph{{Dark Matter Search Results from a One
  Ton-Year Exposure of XENON1T}},
  \href{https://doi.org/10.1103/PhysRevLett.121.111302}{\emph{Phys. Rev. Lett.}
  {\bfseries 121} (2018) 111302}
  [\href{https://arxiv.org/abs/1805.12562}{{\ttfamily 1805.12562}}].

\bibitem{Aprile:2015uzo}
{\scshape XENON} collaboration, \emph{{Physics reach of the XENON1T dark matter
  experiment}},
  \href{https://doi.org/10.1088/1475-7516/2016/04/027}{\emph{JCAP} {\bfseries
  04} (2016) 027} [\href{https://arxiv.org/abs/1512.07501}{{\ttfamily
  1512.07501}}].

\bibitem{Aalbers:2016jon}
{\scshape DARWIN} collaboration, \emph{{DARWIN: towards the ultimate dark
  matter detector}},
  \href{https://doi.org/10.1088/1475-7516/2016/11/017}{\emph{JCAP} {\bfseries
  11} (2016) 017} [\href{https://arxiv.org/abs/1606.07001}{{\ttfamily
  1606.07001}}].

\bibitem{Akerib:2018lyp}
{\scshape LUX-ZEPLIN} collaboration, \emph{{Projected WIMP sensitivity of the
  LUX-ZEPLIN dark matter experiment}},
  \href{https://doi.org/10.1103/PhysRevD.101.052002}{\emph{Phys. Rev. D}
  {\bfseries 101} (2020) 052002}
  [\href{https://arxiv.org/abs/1802.06039}{{\ttfamily 1802.06039}}].

\bibitem{Buttazzo:2013uya}
D.~Buttazzo, G.~Degrassi, P.~P. Giardino, G.~F. Giudice, F.~Sala, A.~Salvio
  et~al., \emph{{Investigating the near-criticality of the Higgs boson}},
  \href{https://doi.org/10.1007/JHEP12(2013)089}{\emph{JHEP} {\bfseries 12}
  (2013) 089} [\href{https://arxiv.org/abs/1307.3536}{{\ttfamily 1307.3536}}].

\bibitem{Hall:2013eko}
L.~J. Hall and Y.~Nomura, \emph{{Grand Unification and Intermediate Scale
  Supersymmetry}}, \href{https://doi.org/10.1007/JHEP02(2014)129}{\emph{JHEP}
  {\bfseries 02} (2014) 129} [\href{https://arxiv.org/abs/1312.6695}{{\ttfamily
  1312.6695}}].

\bibitem{Hall:2014vga}
L.~J. Hall, Y.~Nomura and S.~Shirai, \emph{{Grand Unification, Axion, and
  Inflation in Intermediate Scale Supersymmetry}},
  \href{https://doi.org/10.1007/JHEP06(2014)137}{\emph{JHEP} {\bfseries 06}
  (2014) 137} [\href{https://arxiv.org/abs/1403.8138}{{\ttfamily 1403.8138}}].

\bibitem{Fox:2014moa}
P.~J. Fox, G.~D. Kribs and A.~Martin, \emph{{Split Dirac Supersymmetry: An
  Ultraviolet Completion of Higgsino Dark Matter}},
  \href{https://doi.org/10.1103/PhysRevD.90.075006}{\emph{Phys. Rev. D}
  {\bfseries 90} (2014) 075006}
  [\href{https://arxiv.org/abs/1405.3692}{{\ttfamily 1405.3692}}].

\bibitem{Gogoladze:2007qm}
I.~Gogoladze, N.~Okada and Q.~Shafi, \emph{{Higgs boson mass from gauge-Higgs
  unification}},
  \href{https://doi.org/10.1016/j.physletb.2007.08.082}{\emph{Phys. Lett. B}
  {\bfseries 655} (2007) 257}
  [\href{https://arxiv.org/abs/0705.3035}{{\ttfamily 0705.3035}}].

\bibitem{Redi:2012ad}
M.~Redi and A.~Strumia, \emph{{Axion-Higgs Unification}},
  \href{https://doi.org/10.1007/JHEP11(2012)103}{\emph{JHEP} {\bfseries 11}
  (2012) 103} [\href{https://arxiv.org/abs/1208.6013}{{\ttfamily 1208.6013}}].

\bibitem{Hall:2018let}
L.~J. Hall and K.~Harigaya, \emph{{Implications of Higgs Discovery for the
  Strong CP Problem and Unification}},
  \href{https://doi.org/10.1007/JHEP10(2018)130}{\emph{JHEP} {\bfseries 10}
  (2018) 130} [\href{https://arxiv.org/abs/1803.08119}{{\ttfamily
  1803.08119}}].

\bibitem{Dunsky:2019api}
D.~Dunsky, L.~J. Hall and K.~Harigaya, \emph{{Higgs Parity, Strong CP, and Dark
  Matter}}, \href{https://doi.org/10.1007/JHEP07(2019)016}{\emph{JHEP}
  {\bfseries 07} (2019) 016}
  [\href{https://arxiv.org/abs/1902.07726}{{\ttfamily 1902.07726}}].

\bibitem{Hall:2019qwx}
L.~J. Hall and K.~Harigaya, \emph{{Higgs Parity Grand Unification}},
  \href{https://doi.org/10.1007/JHEP11(2019)033}{\emph{JHEP} {\bfseries 11}
  (2019) 033} [\href{https://arxiv.org/abs/1905.12722}{{\ttfamily
  1905.12722}}].

\bibitem{Dunsky:2019upk}
D.~Dunsky, L.~J. Hall and K.~Harigaya, \emph{{Dark Matter, Dark Radiation and
  Gravitational Waves from Mirror Higgs Parity}},
  \href{https://doi.org/10.1007/JHEP02(2020)078}{\emph{JHEP} {\bfseries 02}
  (2020) 078} [\href{https://arxiv.org/abs/1908.02756}{{\ttfamily
  1908.02756}}].

\bibitem{Dunsky:2020dhn}
D.~Dunsky, L.~J. Hall and K.~Harigaya, \emph{{Sterile Neutrino Dark Matter and
  Leptogenesis in Left-Right Higgs Parity}},
  \href{https://arxiv.org/abs/2007.12711}{{\ttfamily 2007.12711}}.

\bibitem{Seidel:2013sqa}
K.~Seidel, F.~Simon, M.~Tesar and S.~Poss, \emph{{Top quark mass measurements
  at and above threshold at CLIC}},
  \href{https://doi.org/10.1140/epjc/s10052-013-2530-7}{\emph{Eur. Phys. J. C}
  {\bfseries 73} (2013) 2530}
  [\href{https://arxiv.org/abs/1303.3758}{{\ttfamily 1303.3758}}].

\bibitem{Horiguchi:2013wra}
T.~Horiguchi, A.~Ishikawa, T.~Suehara, K.~Fujii, Y.~Sumino, Y.~Kiyo et~al.,
  \emph{{Study of top quark pair production near threshold at the ILC}},
  \href{https://arxiv.org/abs/1310.0563}{{\ttfamily 1310.0563}}.

\bibitem{Kiyo:2015ooa}
Y.~Kiyo, G.~Mishima and Y.~Sumino, \emph{{Strong IR Cancellation in Heavy
  Quarkonium and Precise Top Mass Determination}},
  \href{https://doi.org/10.1007/JHEP11(2015)084}{\emph{JHEP} {\bfseries 11}
  (2015) 084} [\href{https://arxiv.org/abs/1506.06542}{{\ttfamily
  1506.06542}}].

\bibitem{Beneke:2015kwa}
M.~Beneke, Y.~Kiyo, P.~Marquard, A.~Penin, J.~Piclum and M.~Steinhauser,
  \emph{{Next-to-Next-to-Next-to-Leading Order QCD Prediction for the Top
  Antitop $S$-Wave Pair Production Cross Section Near Threshold in $e^+e^-$
  Annihilation}},
  \href{https://doi.org/10.1103/PhysRevLett.115.192001}{\emph{Phys. Rev. Lett.}
  {\bfseries 115} (2015) 192001}
  [\href{https://arxiv.org/abs/1506.06864}{{\ttfamily 1506.06864}}].

\bibitem{Gomez-Ceballos:2013zzn}
{\scshape TLEP Design Study Working Group} collaboration, \emph{{First Look at
  the Physics Case of TLEP}},
  \href{https://doi.org/10.1007/JHEP01(2014)164}{\emph{JHEP} {\bfseries 01}
  (2014) 164} [\href{https://arxiv.org/abs/1308.6176}{{\ttfamily 1308.6176}}].

\bibitem{Lepage:2014fla}
G.~P. Lepage, P.~B. Mackenzie and M.~E. Peskin, \emph{{Expected Precision of
  Higgs Boson Partial Widths within the Standard Model}},
  \href{https://arxiv.org/abs/1404.0319}{{\ttfamily 1404.0319}}.

\bibitem{Cepeda:2019klc}
M.~Cepeda et~al., \emph{{Report from Working Group 2}: {Higgs Physics at the
  HL-LHC and HE-LHC}},
  \href{https://doi.org/10.23731/CYRM-2019-007.221}{\emph{CERN Yellow Rep.
  Monogr.} {\bfseries 7} (2019) 221}
  [\href{https://arxiv.org/abs/1902.00134}{{\ttfamily 1902.00134}}].

\bibitem{Lee:1977ua}
B.~W. Lee and S.~Weinberg, \emph{{Cosmological Lower Bound on Heavy Neutrino
  Masses}}, \href{https://doi.org/10.1103/PhysRevLett.39.165}{\emph{Phys. Rev.
  Lett.} {\bfseries 39} (1977) 165}.

\bibitem{Cirelli:2005uq}
M.~Cirelli, N.~Fornengo and A.~Strumia, \emph{{Minimal dark matter}},
  \href{https://doi.org/10.1016/j.nuclphysb.2006.07.012}{\emph{Nucl. Phys. B}
  {\bfseries 753} (2006) 178}
  [\href{https://arxiv.org/abs/hep-ph/0512090}{{\ttfamily hep-ph/0512090}}].

\bibitem{Giudice:2011cg}
G.~F. Giudice and A.~Strumia, \emph{{Probing High-Scale and Split Supersymmetry
  with Higgs Mass Measurements}},
  \href{https://doi.org/10.1016/j.nuclphysb.2012.01.001}{\emph{Nucl. Phys. B}
  {\bfseries 858} (2012) 63} [\href{https://arxiv.org/abs/1108.6077}{{\ttfamily
  1108.6077}}].

\bibitem{Kurata:2012nf}
Y.~Kurata and N.~Maekawa, \emph{{Averaged Number of the Lightest Supersymmetric
  Particles in Decay of Superheavy Particle with Long Lifetime}},
  \href{https://doi.org/10.1143/PTP.127.657}{\emph{Prog. Theor. Phys.}
  {\bfseries 127} (2012) 657}
  [\href{https://arxiv.org/abs/1201.3696}{{\ttfamily 1201.3696}}].

\bibitem{Harigaya:2016vda}
K.~Harigaya, T.~Lin and H.~K. Lou, \emph{{GUTzilla Dark Matter}},
  \href{https://doi.org/10.1007/JHEP09(2016)014}{\emph{JHEP} {\bfseries 09}
  (2016) 014} [\href{https://arxiv.org/abs/1606.00923}{{\ttfamily
  1606.00923}}].

\bibitem{Kawasaki:1999na}
M.~Kawasaki, K.~Kohri and N.~Sugiyama, \emph{{Cosmological constraints on late
  time entropy production}},
  \href{https://doi.org/10.1103/PhysRevLett.82.4168}{\emph{Phys. Rev. Lett.}
  {\bfseries 82} (1999) 4168}
  [\href{https://arxiv.org/abs/astro-ph/9811437}{{\ttfamily
  astro-ph/9811437}}].

\bibitem{Kawasaki:2000en}
M.~Kawasaki, K.~Kohri and N.~Sugiyama, \emph{{MeV scale reheating temperature
  and thermalization of neutrino background}},
  \href{https://doi.org/10.1103/PhysRevD.62.023506}{\emph{Phys. Rev. D}
  {\bfseries 62} (2000) 023506}
  [\href{https://arxiv.org/abs/astro-ph/0002127}{{\ttfamily
  astro-ph/0002127}}].

\bibitem{Hasegawa:2019jsa}
T.~Hasegawa, N.~Hiroshima, K.~Kohri, R.~S. Hansen, T.~Tram and S.~Hannestad,
  \emph{{MeV-scale reheating temperature and thermalization of oscillating
  neutrinos by radiative and hadronic decays of massive particles}},
  \href{https://doi.org/10.1088/1475-7516/2019/12/012}{\emph{JCAP} {\bfseries
  12} (2019) 012} [\href{https://arxiv.org/abs/1908.10189}{{\ttfamily
  1908.10189}}].

\bibitem{Harigaya:2014waa}
K.~Harigaya, M.~Kawasaki, K.~Mukaida and M.~Yamada, \emph{{Dark Matter
  Production in Late Time Reheating}},
  \href{https://doi.org/10.1103/PhysRevD.89.083532}{\emph{Phys. Rev. D}
  {\bfseries 89} (2014) 083532}
  [\href{https://arxiv.org/abs/1402.2846}{{\ttfamily 1402.2846}}].

\bibitem{Garcia:2018wtq}
M.~A. Garcia and M.~A. Amin, \emph{{Prethermalization production of dark
  matter}}, \href{https://doi.org/10.1103/PhysRevD.98.103504}{\emph{Phys. Rev.
  D} {\bfseries 98} (2018) 103504}
  [\href{https://arxiv.org/abs/1806.01865}{{\ttfamily 1806.01865}}].

\bibitem{Harigaya:2019tzu}
K.~Harigaya, K.~Mukaida and M.~Yamada, \emph{{Dark Matter Production during the
  Thermalization Era}},
  \href{https://doi.org/10.1007/JHEP07(2019)059}{\emph{JHEP} {\bfseries 07}
  (2019) 059} [\href{https://arxiv.org/abs/1901.11027}{{\ttfamily
  1901.11027}}].

\bibitem{Chung:1998rq}
D.~J. Chung, E.~W. Kolb and A.~Riotto, \emph{{Production of massive particles
  during reheating}},
  \href{https://doi.org/10.1103/PhysRevD.60.063504}{\emph{Phys. Rev. D}
  {\bfseries 60} (1999) 063504}
  [\href{https://arxiv.org/abs/hep-ph/9809453}{{\ttfamily hep-ph/9809453}}].

\bibitem{Giudice:2000ex}
G.~F. Giudice, E.~W. Kolb and A.~Riotto, \emph{{Largest temperature of the
  radiation era and its cosmological implications}},
  \href{https://doi.org/10.1103/PhysRevD.64.023508}{\emph{Phys. Rev. D}
  {\bfseries 64} (2001) 023508}
  [\href{https://arxiv.org/abs/hep-ph/0005123}{{\ttfamily hep-ph/0005123}}].

\bibitem{Micha:2002ey}
R.~Micha and I.~I. Tkachev, \emph{{Relativistic turbulence: A Long way from
  preheating to equilibrium}},
  \href{https://doi.org/10.1103/PhysRevLett.90.121301}{\emph{Phys. Rev. Lett.}
  {\bfseries 90} (2003) 121301}
  [\href{https://arxiv.org/abs/hep-ph/0210202}{{\ttfamily hep-ph/0210202}}].

\bibitem{Micha:2004bv}
R.~Micha and I.~I. Tkachev, \emph{{Turbulent thermalization}},
  \href{https://doi.org/10.1103/PhysRevD.70.043538}{\emph{Phys. Rev. D}
  {\bfseries 70} (2004) 043538}
  [\href{https://arxiv.org/abs/hep-ph/0403101}{{\ttfamily hep-ph/0403101}}].

\bibitem{Co:2020xaf}
R.~T. Co, E.~Gonzalez and K.~Harigaya, \emph{{Increasing Temperature toward the
  Completion of Reheating}},
  \href{https://arxiv.org/abs/2007.04328}{{\ttfamily 2007.04328}}.

\bibitem{Fujita:2016vfj}
T.~Fujita and K.~Harigaya, \emph{{Hubble induced mass after inflation in
  spectator field models}},
  \href{https://doi.org/10.1088/1475-7516/2016/12/014}{\emph{JCAP} {\bfseries
  12} (2016) 014} [\href{https://arxiv.org/abs/1607.07058}{{\ttfamily
  1607.07058}}].

\bibitem{Green:1999yh}
A.~M. Green, \emph{{Supersymmetry and primordial black hole abundance
  constraints}}, \href{https://doi.org/10.1103/PhysRevD.60.063516}{\emph{Phys.
  Rev. D} {\bfseries 60} (1999) 063516}
  [\href{https://arxiv.org/abs/astro-ph/9903484}{{\ttfamily
  astro-ph/9903484}}].

\bibitem{Fujita:2014hha}
T.~Fujita, M.~Kawasaki, K.~Harigaya and R.~Matsuda, \emph{{Baryon asymmetry,
  dark matter, and density perturbation from primordial black holes}},
  \href{https://doi.org/10.1103/PhysRevD.89.103501}{\emph{Phys. Rev. D}
  {\bfseries 89} (2014) 103501}
  [\href{https://arxiv.org/abs/1401.1909}{{\ttfamily 1401.1909}}].

\bibitem{Coleman:1977py}
S.~R. Coleman, \emph{{The Fate of the False Vacuum. 1. Semiclassical Theory}},
  \href{https://doi.org/10.1103/PhysRevD.16.1248}{\emph{Phys. Rev. D}
  {\bfseries 15} (1977) 2929}.

\bibitem{Sato:2019wpo}
R.~Sato, \emph{{SimpleBounce : a simple package for the false vacuum decay}},
  \href{https://doi.org/10.1016/j.cpc.2020.107566}{\emph{Comput. Phys. Commun.}
  {\bfseries 258} (2021) 107566}
  [\href{https://arxiv.org/abs/1908.10868}{{\ttfamily 1908.10868}}].

\end{thebibliography}\endgroup

\end{document}